\documentclass[preprint]{aastex62}

\usepackage{bm}
\usepackage{threeparttable}
\usepackage{amsmath,amssymb}
\usepackage{textcomp}
\usepackage{booktabs}
\usepackage{natbib}
\usepackage{chapterbib}
\usepackage{comment}
\usepackage{here}
\usepackage{graphicx}
\usepackage{ulem}

\newcommand{\thisevent}{OGLE-2018-BLG-1185}
\newcommand{\Spitzer}{{\em Spitzer}}
\newcommand{\bdv}[1]{\mbox{\boldmath$#1$}}
\def\bmu{{\bdv\mu}}
\def\bpi{{\bdv\pi}}

\shortauthors{Kondo et al.}

\begin{document}

\title{OGLE-2018-BLG-1185b : A Low-Mass Microlensing Planet Orbiting a Low-Mass Dwarf}

\author{Iona Kondo}
\affiliation{MOA collaboration}
\affiliation{Department of Earth and Space Science, Graduate School of Science, Osaka University, Toyonaka, Osaka 560-0043, Japan}

\author{Jennifer C. Yee}
\affiliation{The \Spitzer Team}
\affiliation{The KMTNet Collaboration}
\affiliation{Center for Astrophysics $|$ Harvard \& Smithsonian, 60 Garden St., Cambridge, MA 02138, USA}

\author{David P.~Bennett}
\affiliation{MOA collaboration}
\affiliation{Code 667, NASA Goddard Space Flight Center, Greenbelt, MD 20771, USA}
\affiliation{Department of Astronomy, University of Maryland, College Park, MD 20742, USA}

\author{Takahiro Sumi}
\affiliation{MOA collaboration}
\affiliation{Department of Earth and Space Science, Graduate School of Science, Osaka University, Toyonaka, Osaka 560-0043, Japan}

\author{Naoki Koshimoto}
\affiliation{MOA collaboration}
\affiliation{Department of Astronomy, Graduate School of Science, The University of Tokyo, 7-3-1 Hongo, Bunkyo-ku, Tokyo 113-0033, Japan}

\author{Ian A. Bond}
\affiliation{MOA collaboration}
\affiliation{Institute of Natural and Mathematical Sciences, Massey University, Auckland 0745, New Zealand}

\author{Andrew Gould}
\affiliation{The \Spitzer Team}
\affiliation{The KMTNet Collaboration}
\affiliation{Max-Planck-Institute for Astronomy, Konigstuhl 17, 69117 Heidelberg, Germany}
\affiliation{Department of Astronomy Ohio State University, 140 W. 18th Ave., Columbus, OH 43210, USA}

\author{Andrzej Udalski}
\affiliation{OGLE collaboration}
\affiliation{Warsaw University Observatory, Al. Ujazdowskie 4, 00-478 Warszawa, Poland}

\author{Yossi Shvartzvald}
\affiliation{The \Spitzer Team}
\affiliation{The KMTNet Collaboration}
\affiliation{Department of Particle Physics and Astrophysics, Weizmann Institute of Science, Rehovot 76100, Israel}

\author{Youn Kil Jung}
\affiliation{The KMTNet Collaboration}
\affiliation{Korea Astronomy and Space Science Institute, Daejon 34055, Republic of Korea}
\affiliation{University of Science and Technology, Korea, (UST), 217 Gajeong-ro Yuseong-gu, Daejeon 34113, Republic of Korea}

\author{Weicheng Zang}
\affiliation{The \Spitzer Team}
\affiliation{The KMTNet Collaboration}
\affiliation{Department of Astronomy, Tsinghua University, Beijing 100084, China}

\author{Valerio Bozza}
\affiliation{MiNDSTEp Collaboration}
\affiliation{The ROME/REA project team}
\affiliation{Dipartimento di Fisica “E.R. Caianiello”, Universit\`{a} di Salerno, Via Giovanni Paolo II 132, 84084, Fisciano, Italy}
\affiliation{Istituto Nazionale di Fisica Nucleare, Sezione di Napoli, Napoli, Italy}

\author{Etienne Bachelet}
\affiliation{The ROME/REA project team}
\affiliation{MiNDSTEp Collaboration}
\affiliation{Las Cumbres Observatory, 6740 Cortona Drive, Suite 102,93117 Goleta, CA, USA}

\author{Markus P.G. Hundertmark}
\affiliation{The ROME/REA project team}
\affiliation{MiNDSTEp Collaboration}
\affiliation{Astronomisches Rechen-Institut, Zentrum f\"{u}r Astronomie der Universit\"{a}t Heidelberg (ZAH), 69120 Heidelberg, Germany}

\author{Nicholas J. Rattenbury}
\affiliation{MOA collaboration}
\affiliation{Department of Physics, University of Auckland, Private Bag 92019, Auckland, New Zealand}

\collaboration{(Leading Authors)}

\author{F. Abe}
\affiliation{Institute for Space-Earth Environmental Research, Nagoya University, Nagoya 464-8601, Japan}
\author{R. Barry}
\affiliation{Code 667, NASA Goddard Space Flight Center, Greenbelt, MD 20771, USA}
\author{A. Bhattacharya}
\affiliation{Code 667, NASA Goddard Space Flight Center, Greenbelt, MD 20771, USA}
\affiliation{Department of Astronomy, University of Maryland, College Park, MD 20742, USA}
\author{M. Donachie}
\affiliation{Department of Physics, University of Auckland, Private Bag 92019, Auckland, New Zealand}
\author{A. Fukui}
\affiliation{Department of Earth and Planetary Science, Graduate School of Science, The University of Tokyo, 7-3-1 Hongo, Bunkyo-ku, Tokyo 113-0033, Japan}
\affiliation{Instituto de Astrof\'isica de Canarias, V\'ia L\'actea s/n, E-38205 La Laguna, Tenerife, Spain}
\author{H. Fujii}
\affiliation{Institute for Space-Earth Environmental Research, Nagoya University, Nagoya 464-8601, Japan}
\author{Y. Hirao}
\affiliation{Department of Earth and Space Science, Graduate School of Science, Osaka University, Toyonaka, Osaka 560-0043, Japan}
\author{S. Ishitani Silva}
\affiliation{Department of Physics, The Catholic University of America, Washington, DC 20064, USA}
\affiliation{Code 667, NASA Goddard Space Flight Center, Greenbelt, MD 20771, USA}
\author{Y. Itow}
\affiliation{Institute for Space-Earth Environmental Research, Nagoya University, Nagoya 464-8601, Japan}
\author{R. Kirikawa}
\affiliation{Department of Earth and Space Science, Graduate School of Science, Osaka University, Toyonaka, Osaka 560-0043, Japan}
\author{M. C. A. Li}
\affiliation{Department of Physics, University of Auckland, Private Bag 92019, Auckland, New Zealand}
\author{Y. Matsubara}
\affiliation{Institute for Space-Earth Environmental Research, Nagoya University, Nagoya 464-8601, Japan}
\author{S. Miyazaki}
\affiliation{Department of Earth and Space Science, Graduate School of Science, Osaka University, Toyonaka, Osaka 560-0043, Japan}
\author{Y. Muraki}
\affiliation{Institute for Space-Earth Environmental Research, Nagoya University, Nagoya 464-8601, Japan}
\author{G. Olmschenk}
\affiliation{Code 667, NASA Goddard Space Flight Center, Greenbelt, MD 20771, USA}
\author{C. Ranc}
\affiliation{Sorbonne Universit\'e, CNRS, UMR 7095, Institut d'Astrophysique de Paris, 98 bis bd Arago, 75014 Paris, France}
\author{Y. Satoh}
\affiliation{Department of Earth and Space Science, Graduate School of Science, Osaka University, Toyonaka, Osaka 560-0043, Japan}
\author{H. Shoji}
\affiliation{Department of Earth and Space Science, Graduate School of Science, Osaka University, Toyonaka, Osaka 560-0043, Japan}
\author{D. Suzuki}
\affiliation{Department of Earth and Space Science, Graduate School of Science, Osaka University, Toyonaka, Osaka 560-0043, Japan}
\author{Y. Tanaka}
\affiliation{Department of Earth and Space Science, Graduate School of Science, Osaka University, Toyonaka, Osaka 560-0043, Japan}
\author{P. J. Tristram}
\affiliation{University of Canterbury Mt.\ John Observatory, P.O. Box 56, Lake Tekapo 8770, New Zealand}
\author{T. Yamawaki}
\affiliation{Department of Earth and Space Science, Graduate School of Science, Osaka University, Toyonaka, Osaka 560-0043, Japan}
\author{A. Yonehara}
\affiliation{Department of Physics, Faculty of Science, Kyoto Sangyo University, 603-8555 Kyoto, Japan}
\collaboration{(The MOA collaboration)}


\author{P. Mr{\'o}z}
\affiliation{Division of Physics, Mathematics, and Astronomy, California Institute of Technology, Pasadena, CA 91125, USA}
\author{R. Poleski}
\affiliation{Warsaw University Observatory, Al. Ujazdowskie 4, 00-478 Warszawa, Poland}
\author{J. Skowron}
\affiliation{Warsaw University Observatory, Al. Ujazdowskie 4, 00-478 Warszawa, Poland}
\author{M.~K. Szyma{\'n}ski}
\affiliation{Warsaw University Observatory, Al. Ujazdowskie 4, 00-478 Warszawa, Poland}
\author{I. Soszy{\'n}ski}
\affiliation{Warsaw University Observatory, Al. Ujazdowskie 4, 00-478 Warszawa, Poland}
\author{S. Koz{\l}owski}
\affiliation{Warsaw University Observatory, Al. Ujazdowskie 4, 00-478 Warszawa, Poland}
\author{P. Pietrukowicz}
\affiliation{Warsaw University Observatory, Al. Ujazdowskie 4, 00-478 Warszawa, Poland}
\author{K. Ulaczyk}
\affiliation{Warsaw University Observatory, Al. Ujazdowskie 4, 00-478 Warszawa, Poland}
\affiliation{Department of Physics, University of Warwick, Gibbet Hill Road, Coventry, CV4 7AL, UK}
\author{K. A. Rybicki}
\affiliation{Warsaw University Observatory, Al. Ujazdowskie 4, 00-478 Warszawa, Poland}
\author{P. Iwanek}
\affiliation{Warsaw University Observatory, Al. Ujazdowskie 4, 00-478 Warszawa, Poland}
\author{M. Wrona}
\affiliation{Warsaw University Observatory, Al. Ujazdowskie 4, 00-478 Warszawa, Poland}
\collaboration{(The OGLE collaboration)}

\author{M. D. Albrow}
\affiliation{University of Canterbury, Department of Physics and Astronomy, Private Bag 4800, Christchurch 8020, New Zealand}
\author{S.-J. Chung}
\affiliation{Korea Astronomy and Space Science Institute, Daejon 34055, Republic of Korea}
\affiliation{University of Science and Technology, Korea, (UST), 217 Gajeong-ro Yuseong-gu, Daejeon 34113, Republic of Korea}
\author{C. Han}
\affiliation{Department of Physics, Chungbuk National University, Cheongju 28644, Republic of Korea}
\author{K.-H. Hwang}
\affiliation{Korea Astronomy and Space Science Institute, Daejon 34055, Republic of Korea}
\author{H.-W. Kim}
\affiliation{Korea Astronomy and Space Science Institute, Daejon 34055, Republic of Korea}
\author{I.-G. Shin}
\affiliation{Korea Astronomy and Space Science Institute, Daejon 34055, Republic of Korea}
\author{S.-M. Cha}
\affiliation{Korea Astronomy and Space Science Institute, Daejon 34055, Republic of Korea}
\affiliation{School of Space Research, Kyung Hee University, Yongin, Kyeonggi 17104, Republic of Korea}
\author{D.-J. Kim}
\affiliation{Korea Astronomy and Space Science Institute, Daejon 34055, Republic of Korea}
\author{S.-L. Kim}
\affiliation{Korea Astronomy and Space Science Institute, Daejon 34055, Republic of Korea}
\affiliation{University of Science and Technology, Korea, (UST), 217 Gajeong-ro Yuseong-gu, Daejeon 34113, Republic of Korea}
\author{C.-U. Lee}
\affiliation{Korea Astronomy and Space Science Institute, Daejon 34055, Republic of Korea}
\author{D.-J. Lee}
\affiliation{Korea Astronomy and Space Science Institute, Daejon 34055, Republic of Korea}
\author{Y. Lee}
\affiliation{Korea Astronomy and Space Science Institute, Daejon 34055, Republic of Korea}
\affiliation{School of Space Research, Kyung Hee University, Yongin, Kyeonggi 17104, Republic of Korea}
\author{B.-G. Park}
\affiliation{Korea Astronomy and Space Science Institute, Daejon 34055, Republic of Korea}
\affiliation{University of Science and Technology, Korea, (UST), 217 Gajeong-ro Yuseong-gu, Daejeon 34113, Republic of Korea}
\author{R. W. Pogge}
\affiliation{Department of Astronomy, Ohio State University, 140 W. 18th Ave., Columbus, OH 43210, USA}
\author{Y.-H. Ryu}
\affiliation{Korea Astronomy and Space Science Institute, Daejon 34055, Republic of Korea}
\collaboration{(The KMTNet collaboration)}

\author{C. A. Beichman}
\affiliation{IPAC, Mail Code 100-22, Caltech, 1200 E. California Blvd., Pasadena, CA 91125, USA}
\author{G. Bryden}
\affiliation{Jet Propulsion Laboratory, California Institute of Technology, 4800 Oak Grove Drive, Pasadena, CA 91109, USA}
\author{S. Calchi Novati}
\affiliation{IPAC, Mail Code 100-22, Caltech, 1200 E. California Blvd., Pasadena, CA 91125, USA}
\author{S. Carey}
\affiliation{IPAC, Mail Code 100-22, Caltech, 1200 E. California Blvd., Pasadena, CA 91125, USA}
\author{B. S. Gaudi}
\affiliation{Department of Astronomy, The Ohio State University, 140 W. 18th Ave., Columbus, OH 43210, USA}
\author{C. B. Henderson}
\affiliation{IPAC, Mail Code 100-22, Caltech, 1200 E. California Blvd., Pasadena, CA 91125, USA}
\author{W. Zhu}
\affiliation{Canadian Institute for Theoretical Astrophysics, University of Toronto, 60 St George Street, Toronto, ON M5S 3H8, Canada}
\collaboration{(The $Spitzer$ team)}

\author{D. Maoz}
\affiliation{School of Physics and Astronomy, Tel-Aviv University, Tel-Aviv 6997801, Israel}
\author{M. T. Penny}
\affiliation{Department of Physics \& Astronomy, Louisiana State University, Baton Rouge, LA 70803-4001}

\collaboration{(The LCO Follow-up Team)}

\author{M. Dominik}
\affiliation{University of St Andrews, Centre for Exoplanet Science, SUPA School of Physics \& Astronomy, North Haugh, St Andrews, KY16 9SS, United Kingdom}
\author{U. G. J{\o}rgensen}
\affiliation{Centre for ExoLife Sciences, Niels Bohr Institute, University of Copenhagen {\O}ster Voldgade 5, 1350 - Copenhagen, Denmark}
\author{P. Longa-Pe{\~{ n}}a}
\affiliation{Centro de Astronomía (CITEVA), Universidad de Antofagasta, Avda. U. de Antofagasta 02800, Antofagasta, Chile}
\author{N. Peixinho}
\affiliation{Univ Coimbra, Instituto de Astrof\'{\i}sica e Ci\^encias do Espa\c co, OGAUC, R. do Observat\'orio s/n, 3040-004 Coimbra, Portugal}
\author{S. Sajadian}
\affiliation{Department of Physics, Isfahan University of Technology, Isfahan, Iran}
\author{J. Skottfelt}
\affiliation{Centre for Electronic Imaging, Department of Physical Sciences, The Open University, Milton Keynes, MK7 6AA, UK}
\author{C. Snodgrass}
\affiliation{Institute for Astronomy, University of Edinburgh, Royal Observatory, Edinburgh EH9 3HJ, UK}
\author{J. Tregloan-Reed}
\affiliation{Instituto de Investigación en Astronomia y Ciencias Planetarias, Universidad de Atacama, Copiap\'o, Atacama, Chile}
\author{M. J. Burgdorf}
\affiliation{Universit\"{a}t Hamburg, Faculty of Mathematics, Informatics and Natural Sciences, Department of Earth Sciences, Meteorological Institute, Bundesstra{\ss}e 55, 20146 Hamburg, Germany}
\author{J. Campbell-White}
\affiliation{SUPA, School of Science and Engineering, University of Dundee, Nethergate, Dundee DD1 4HN, UK}
\author{S. Dib}
\affiliation{Centre for ExoLife Sciences, Niels Bohr Institute, University of Copenhagen {\O}ster Voldgade 5, 1350 - Copenhagen, Denmark}
\affiliation{Max Planck Institute for Astronomy, K\"{o}nigstuhl 17, 69117, Heidelberg, Germany}
\author{Y. I. Fujii}
\affiliation{Institute for Advanced Research, Nagoya University, Furo-cho, Chikusa-ku, Nagoya, 464-8601, Japan}
\affiliation{Department of Physics, Nagoya University, Furo-cho, Chikusa-ku, Nagoya 464-8602, Japan}
\affiliation{Niels Bohr Institute \& Centre for Star and Planet Formation, University of Copenhagen {\O}ster Voldgade 5, 1350 - Copenhagen, Denmark}
\affiliation{Centre for ExoLife Sciences, Niels Bohr Institute, University of Copenhagen {\O}ster Voldgade 5, 1350 - Copenhagen, Denmark}
\affiliation{Graduate School of Human and Environmental Studies, Kyoto University, Yoshida-Nihonmatsu, Sakyo, Kyoto 606-8501, Japan}
\author{T. C. Hinse}
\affiliation{Institute of Astronomy, Faculty of Physics, Astronomy and Informatics, Nicolaus Copernicus University, Grudziadzka 5, 87-100 Torun, Poland}
\affiliation{Chungnam National University, Department of Astronomy and Space Science, 34134 Daejeon, Republic of Korea}
\author{E. Khalouei}
\affiliation{Department of Physics, Sharif University of Technology, P. O. Box 11365-9161, Tehran, Iran}
\author{S. Rahvar}
\affiliation{Department of Physics, Sharif University of Technology, P. O. Box 11365-9161, Tehran, Iran}
\author{M. Rabus}
\affiliation{Las Cumbres Observatory Global Telescope, 6740 Cortona Dr, Suite 102, Goleta, CA 93111, USA}
\affiliation{Department of Physics, University of California, Santa Barbara, CA 93106-9530, USA}
\author{J. Southworth}
\affiliation{Astrophysics Group, Keele University, Staffordshire, ST5 5BG, UK}
\collaboration{(The MindSTEp Collaboration)}

\author{Y. Tsapras}
\affiliation{Zentrum f{\"u}r Astronomie der Universit{\"a}t Heidelberg, Astronomisches Rechen-Institut, M{\"o}nchhofstr. 12-14, 69120 Heidelberg, Germany}
\author{R. A. Street}
\affiliation{Las Cumbres Observatory Global Telescope Network, 6740 Cortona Drive, suite 102, Goleta, CA 93117, USA}
\author{D. M. Bramich}
\affiliation{Center for Space Science, NYUAD Institute, New York University Abu Dhabi, PO Box 129188, Saadiyat Island, Abu Dhabi, UAE.}
\affiliation{Division of Science, New York University Abu Dhabi, PO Box 129188, Saadiyat Island, Abu Dhabi, UAE.}
\affiliation{Division of Engineering, New York University Abu Dhabi, PO Box 129188, Saadiyat Island, Abu Dhabi, UAE.}
\author{A. Cassan}
\affiliation{Institut d’Astrophysique de Paris, Sorbonne Universit{\'e}, CNRS, UMR 7095, 98 bis bd Arago, 75014 Paris, France }
\author{K. Horne}
\affiliation{SUPA, School of Physics \& Astronomy, University of St Andrews, North Haugh, St Andrews KY16 9SS, UK}
\author{J. Wambsganss}
\affiliation{Zentrum f{\"u}r Astronomie der Universit{\"a}t Heidelberg, Astronomisches Rechen-Institut, M{\"o}nchhofstr. 12-14, 69120 Heidelberg, Germany}
\author{S. Mao}
\affiliation{National Astronomical Observatories, Chinese Academy of Sciences, 100012 Beijing, China}
\author{A. Saha}
\affiliation{National Optical Astronomy Observatory, 950 North Cherry Ave., Tucson, AZ 85719, USA}
\collaboration{(The ROME/REA project team)}

\begin{abstract}
We report the analysis of planetary microlensing event OGLE-2018-BLG-1185, which was observed by a large number of ground-based telescopes and by the $Spitzer$ Space Telescope. The ground-based light curve indicates a low planet-host star mass ratio of $q = (6.9 \pm 0.2) \times 10^{-5}$, which is near the peak of the wide-orbit exoplanet mass-ratio distribution. We estimate the host star and planet masses with a Bayesian analysis using the measured angular Einstein radius under the assumption that stars of all masses have an equal probability to host this planet. The flux variation observed by $Spitzer$ was marginal, but still places a constraint on the microlens parallax. Imposing a conservative constraint that this flux variation should be $\Delta f_{\rm Spz} < 4$ instrumental flux units
indicates a host mass of $M_{\rm host} =  0.37^{+0.35}_{-0.21}\  M_\odot$ and a planet mass of $m_{\rm p} =  8.4^{+7.9}_{-4.7}\ M_\oplus$. A Bayesian analysis including the full parallax constraint from \Spitzer\ suggests smaller host star and planet masses of $M_{\rm host} =  0.091^{+0.064}_{-0.018}\ M_\odot$ and $m_{\rm p} = 2.1^{+1.5}_{-0.4}\ M_\oplus$, respectively.
Future high-resolution imaging observations with {\it HST} or ELTs could distinguish between these two scenarios and help to reveal the planetary system properties in more detail.

\end{abstract}


\section{Introduction}
\label{sec-intr}

The gravitational microlensing method has a unique sensitivity to low-mass planets \citep{1996ApJ...472..660B} 
beyond the snow line of the host star \citep{1992ApJ...396..104G}, where the core accretion theory predicts that planet formation is most efficient \citep{1993ARA&A..31..129L, 1996Icar..124...62P}.
The Microlensing Observations in Astrophysics (MOA; \citealp{2001MNRAS.327..868B,2003ApJ...591..204S}) collaboration presented the most complete statistical analysis of planets found by microlensing to date and the best measurement of the planet distribution beyond the snow line in \citet{2016ApJ...833..145S}. 
They found that the mass-ratio distribution from the $2007-2012$ MOA-II microlensing survey combined with earlier samples \citep{gould10,cassan12} is well fitted by a broken power-law model. 

Their result shows 
the mass-ratio distribution peaks at $q_{\rm br} = (6.7^{+9.0}_{-1.8})\times 10^{-5}$ with power-law slopes of $n = -0.85^{+0.12}_{-0.13}$ and $p = 2.6^{+4.2}_{-2.1}$ above and below $q_{\rm br}$, respectively\footnote{These values are the median and 68\% confidence level by the Markov Chain Monte Carlo analysis with the thirty planet sample, which are given in Table 5 of \citet{2016ApJ...833..145S}. So the 1$\sigma$ range of the mass-ratio distribution peaks is roughly $q_{\rm br} \sim (0.5-2)\times10^{-4}$. At the same time, they also show that the best fitting parameters are $q_{\rm br} = 1.65\times10^{-4}$ with power-law slopes of $n = -0.92$ and $p = 0.47$ in Table 4 of \citet{2016ApJ...833..145S}.}.
This result is consistent with previous microlensing analyses which suggest that Neptune mass-ratio planets are more common than larger gas giants \citep{2006ApJ...644L..37G,2010ApJ...710.1641S} and further indicates that Neptune mass-ratio planets are, in fact, the most common type of planet (large or small) in wide orbits.

Additionally, \citet{2018ApJ...869L..34S} reveals a disagreement between the measured mass ratio distribution in \citet{2016ApJ...833..145S} and the predictions of the runaway gas accretion scenario \citep{idalin04}, 
which is part of the standard core accretion theory. 
Population synthesis models based on core accretion, including runaway gas accretion, 
predict too few planets in the mass range of approximately 
$20-80 M_\oplus$ compared to those inferred from microlensing observations.
Similar tension is indicated by ALMA observations. \citet{2019MNRAS.488L..12N} compared wide 
orbit (9-99 au) planet candidates with masses of $0.01 M_{\rm Jup}$ to a few $M_{\rm Jup}$ suggested by ALMA proto-planetary disk observations to a population synthesis prediction from the runaway gas accretion scenario.
They found that the scenario predicts fewer sub-Jovian planets than those inferred from the ALMA observation.
3D hydrodynamical simulations of proto-planetary disks do not support the
runaway gas accretion scenario either \citep{2019A&A...630A..82L}.

The peak position of the mass-ratio function and its slope at small mass ratios are uncertain due to the lack of planets with
mass ratios of $q < 5.8 \times 10^{-5}$ in the \citet{2016ApJ...833..145S} sample.
\citet{2018AcA....68....1U} and \citet{2019AJ....157...72J} used 
samples of published planets to refine the estimates of the peak and the low mass-ratio slope of the mass-ratio function.
\citet{2018AcA....68....1U} confirmed the turnover shown in \citet{2016ApJ...833..145S} and obtained the slope index on the low-mass regime, $p\sim 0.73$, using seven published planets with $q<1 \times 10^{-4}$.
\citet{2019AJ....157...72J} found $q_{\rm br}\simeq 0.55 \times 10^{-4}$ using 15 published planets with
low mass ratio ($q<3 \times 10^{-4}$). 
The \citet{2019AJ....157...72J} study was subject to ``publication bias". That is, the planets were not part of a well-defined statistical sample. Instead, these planets were selected for publication for reasons that are not well characterized. Nevertheless, the authors make the case that this publication bias should not be large enough to invalidate their results.   By contrast, the \citet{2018AcA....68....1U} study
only made the implicit assumption that all planets with $q<1\times 10^{-4}$ (and greater than that of the actual
published planet) would have been published.  If this is true (which is very likely),
the study is not subject to publication bias.

A more definitive improvement of the \citet{2016ApJ...833..145S} mass-ratio function can be obtained with an
extension of the MOA-II statistical sample to include additional microlensing seasons (Suzuki et al., in preparation).
The low mass-ratio planet analyzed in this paper, OGLE-2018-BLG-1185Lb,
will be part of that extended sample, and it will contribute to an improved characterization of the low end of the 
wide orbit exoplanet mass-ratio function.

The statistical analysis of the wide-orbit planet population can also be improved
by including information on the lens physical parameters, such as the lens mass, $M_{\rm L}$, and the distance to the lens star, $D_{\rm L}$.
While the lens planet-host mass ratios, $q$, are usually well constrained from the light-curve modeling, we need at least two mass-distance relations in order to derive $M_{\rm L}$ and $D_{\rm L}$ directly.
There are three observables that can yield mass-distance relations: finite source effects, microlens parallax effects, and direct detection of the lens flux. 

In recent years, lens flux detection by high-resolution imaging follow-up observations (such as by the {\em Hubble} Space Telescope ({\it HST}) or Keck) has been done for several microlens planetary systems after the lens and the source are separated enough to be detected \citep{2006ApJ...647L.171B, 2007ApJ...660..781B, 2015ApJ...808..169B, 2020AJ....159...68B, 2014ApJ...780...54B, 2015ApJ...808..170B, 2017AJ....154...59B, 2018AJ....156..289B, 2017AJ....154....3K, 2020AJ....160..121V}. 
However, the required separation for resolving the lens and source depends on their relative brightnesses, and even if they are comparable in brightness, it typically takes a few years for them to separate sufficiently.

If both the Einstein radius $\theta_{\rm E}$ from the finite source effect and the microlens parallax $\pi_{\rm E}$ from the parallax effect are measured, 
we can derive two mass-distance relations as follows,
\begin{equation}
M_{\rm L} = \frac{c^2}{4G} {\theta_{\rm E}}^2 \frac{D_{\rm S}D_{\rm L}}{D_{\rm S}-D_{\rm L}} = \frac{c^2}{4G} \frac{{\rm au}}{{\pi_{\rm E}}^2} \frac{D_{\rm S}-D_{\rm L}}{D_{\rm S}D_{\rm L}},
\label{eqn:ML0}
\end{equation}
where $D_{\rm S}$ is the distance to the source \citep{1992ApJ...392..442G, 2000ApJ...542..785G}.
Finite source effects are detected in most planetary-lens events through the observation of a caustic crossing or a close approach to a caustic cusp, thus enabling the measurement of $\theta_{\rm E}$.

The most common method for measuring the microlens parallax has been via the effects of the motion of the observer, which is called the ``orbital parallax effect.''  In order to detect the orbital parallax, the ratio of $t_{\rm E}$  (typically  $t_{\rm E}$ is $\sim 30$ days) to Earth's orbital period (365 days) should be significant. 
Thus, we only measure the orbital parallax effect for microlensing events with long durations and/or with relatively nearby lens systems, yielding mass measurements in less than half of published microlensing planetary systems.

The most effective method for routinely obtaining a microlens parallax measurement is via  the ``satellite parallax effect'' \citep{1966MNRAS.134..315R}, which is caused by the separation between two observers. 
Because the typical Einstein radius projected onto the observer plane, $\tilde{r}_{\rm E}$, is about 10 au, the satellite parallax effect can be measured for a wide range of microlenses provided the separation between Earth and the satellite is about 1 au (as was the case for \Spitzer).

For the purpose of measuring the Galactic distribution of planets and making mass measurements through the satellite parallax effect, the $Spitzer$ microlensing campaign was carried out from 2014--2019 \citep{2013ApJ...764..107G,2014sptz.prop11006G,2015sptz.prop12013G,2015sptz.prop12015G,2016sptz.prop13005G,2018sptz.prop14012G}. During the six-year program, close to 1000 microlensing events were simultaneously observed from the ground and by $Spitzer$, and there are 11 published\footnote{In addition \citet{Yee21} have submitted a paper on OGLE-2019-BLG-0960.} planets with satellite parallax measurements from $Spitzer$: OGLE-2014-BLG-0124Lb \citep{2015ApJ...799..237U}, OGLE-2015-BLG-0966Lb \citep{2016ApJ...819...93S}, OGLE-2016-BLG-1067Lb \citep{2019AJ....157..121C},  OGLE-2016-BLG-1195Lb \citep{2017ApJ...840L...3S}, OGLE-2016-BLG-1190Lb \citep{2018AJ....155...40R}, OGLE-2017-BLG-1140Lb \citep{2018AJ....155..261C},  TCP J05074264+2447555 \citep{2018MNRAS.476.2962N, 2019AJ....158..206F, 2020ApJ...897..180Z}, OGLE-2018-BLG- 0596Lb \citep{2019AJ....158...28J}, KMT-2018-BLG-0029Lb \citep{2020JKAS...53....9G}, OGLE-2017-BLG-0406Lb \citep{2020AJ....160...74H}, and OGLE-2018-BLG-0799Lb \citep{2020arXiv201008732Z}. Comparing the planet frequency in the disk to that in the bulge could probe the effects of the different environments on the planet formation process.

Obvious correlated noise in the $Spitzer$ photometry was first noted by \citet{2016ApJ...823...63P} and \citet{2017AJ....154..210Z}, but those works did not expect the systematic errors would have a significant effect on the parallax measurements. Indeed, two comparisons of small, heterogeneous samples of published $Spitzer$ microlensing events confirmed this expectation \citep{2019ApJ...873...30S,2020ApJ...891....3Z}. However, 
a larger study \citep{2020AJ....160..177K} of the 50-event statistical sample of \citet{2017AJ....154..210Z} indicated a conflict between the $Spitzer$ microlensing parallax measurements and Galactic models. They suggested that this conflict was probably caused by systematic errors in the $Spitzer$ photometry.
   Based, in part, on the \citet{2020AJ....160..177K} analysis, the $Spitzer$ microlensing team has made a greater effort to understand these systematic errors, including obtaining baseline data in 2019 for many of the earlier planetary events. These additional baseline data proved very useful in characterizing systematics in the $Spitzer$ photometry for three previously published events \citep{2020JKAS...53....9G,2020AJ....160...74H,2020arXiv201008732Z}. Those analyses show that systematics in the \Spitzer\ photometry can be present at the level of 1--2 instrumental flux units, so observed signals in the \Spitzer\ photometry on those scales should be interpreted with caution.

In this paper, we present the analysis of planetary microlensing event OGLE-2018-BLG-1185, which was simultaneously observed by many ground-based telescopes and by the $Spitzer$ Space Telescope. From the ground-based light-curve analysis, the planet-host star mass ratio turns out to be very low, $q\sim 6.9\times10^{-5}$, which is thought to be 
near the peak of the wide-orbit exoplanet mass-ratio distribution in \citet{2016ApJ...833..145S}, \citet{2018AcA....68....1U}, and \citet{2019AJ....157...72J}.
Section \ref{sec-obs} explains the observations and the data reductions. 
Our ground-based light-curve modeling method and results are shown in Section \ref{sec-analysis}. 
In Section \ref{sec-cmd}, we derive the angular Einstein radius from the source magnitude and color and the finite source effect in order to constrain the physical parameters of the planetary system.
In Section \ref{sec-bay}, we estimate the physical properties such as the host star and planet masses based on the ground-based light curve alone by performing a Bayesian analysis using the measured angular Einstein radius under the assumption that stars of all masses have an equal probability to host this planet. We present our parallax analysis including the $Spitzer$ data in Section \ref{sec-sp}.
Finally, we discuss the analysis and summarize our conclusions in Section \ref{sec-dis}.
\\

\section{Observations and Data Reductions}
\label{sec-obs} 
\subsection{Ground-based Survey Observations}

The microlensing event OGLE-2018-BLG-1185 was first discovered on 2018 July 7 (${\rm HJD'}$ $=$ ${\rm HJD} - 2450000$ $\sim$ $8306$), at the J2000 equatorial coordinates $({\rm RA, Dec})$ = ($17^{h}$ $59^{m}$ $10^{s}$.$26, -27^{\circ}$ $50^{\prime}$ $06^{\prime \prime}.3$) 
corresponding to Galactic coordinates $(l, b)$ = $(2.465^{\circ}, -2.004^{\circ})$ by the Optical Gravitational Lensing Experiment (OGLE; \citealp{2003AcA....53..291U}) collaboration. 
The OGLE collaboration conducts a microlensing survey using the 1.3m Warsaw telescope with a 1.4 ${\rm deg^{2}}$  field-of-view (FOV) CCD camera at Las Campanas Observatory in Chile and distributes alerts of the discovery of the microlensing events by the OGLE-IV Early Warning System \citep{ews1,2003AcA....53..291U}. The event is located in the OGLE-IV field BLG 504, which is observed with a cadence of one observation per hour.

The event was also discovered independently on 2018 July 9 by 
the MOA collaboration and identified as MOA-2018-BLG-228 by the MOA alert system \citep{2001MNRAS.327..868B}.
The MOA collaboration conducts a microlensing exoplanet survey toward the Galactic bulge  using the 1.8m MOA-II telescope with a 2.2 ${\rm deg^{2}}$ wide FOV CCD-camera, MOA-cam3 \citep{2008ExA....22...51S} at the University of Canterbury's Mt.\ John Observatory in New Zealand. The MOA survey uses a custom wide-band filter referred to as $R_{\rm MOA}$, 
corresponding to the sum of the Cousins $R$- and $I$-bands and also uses a Johnson $V$-band filter.
The event is located in the MOA field gb10, which is observed at a high cadence of one observation every 15 minutes.

The Korea Microlensing Telescope Network (KMTNet; \citealt{kmtnet}) collaboration conducts a microlensing survey using three 1.6m telescopes each with a 4.0 ${\rm deg^{2}}$ FOV CCD camera. The telescopes are located at Cerro Tololo Interamerican Observatory in Chile (KMTC), South African Astronomical Observatory in South Africa (KMTS), and Siding Spring Observatory in Australia (KMTA). 
This event is located in an overlapping region between two fields (KMTNet BLG03 and BLG43) and was identified by the KMTNet EventFinder \citep{KimKim18_EF} as KMT-2018-BLG-1024.

\subsection{Spitzer Observations}
In order to construct statistical samples from the $Spitzer$ microlensing campaign, \citet{2015ApJ...810..155Y} established detailed protocols for the selection and observational cadence of $Spitzer$ microlensing targets. 
On 2018 July 8 (${\rm HJD'} \sim 8308.25$),
OGLE-2018-BLG-1185 was selected as a ``Subjective, Immediate" (SI) target to be observed with the ``objective" cadence by the $Spitzer$ microlensing team. The selection as SI meant that this event was observed even though it never met the objective criteria established in \citet{2015ApJ...810..155Y}.
The $Spitzer$ Space Telescope began to observe this event on 2018 July 14 (${\rm HJD'} \sim 8313.83$), which was three days after the peak observed from the ground-based telescopes. The ``objective" cadence resulted in approximately one observation per day for the remainder of the observing window (27 days total). These observations were taken with the IRAC camera in the $3.6$ $\mu$m ($L$) band.

\subsection{Ground-based Follow-up Observations}
After the event was selected for \Spitzer\ observations, some ground-based follow-up observations were conducted. 
Microlensing Network for the Detection of Small Terrestrial Exoplanets (MiNDSTEp) used the 1.54m Danish Telescope at La Silla Observatory in Chile and the 0.6m telescope at Salerno University Observatory in Italy.
The Microlensing Follow Up Network ($\mu$FUN) used the 1.3m SMARTS telescope at CTIO in Chile.
Las Cumbres Observatory (LCO; \citealp{2013PASP..125.1031B}) used the 1.0m telescopes at CTIO in Chile, at SSO in Australia, and at SAAO in South Africa, as a part of LCO-$Spitzer$ program.
The ROME/REA team \citep{2019PASP..131l4401T} also used the 1.0m LCO robotic telescopes at CTIO in Chile, at SSO in Australia, and at SAAO in South Africa. A summary of observations from each telescope is given in Table \ref{Table_tel}.

\subsection{Data Reduction}
The OGLE, MOA, and KMTNet data were reduced using the OGLE Difference Image Analysis (DIA) photometry pipeline \citep{2003AcA....53..291U}, the MOA DIA photometry pipeline \citep{2001MNRAS.327..868B}, and the KMTNet pySIS photometry pipeline \citep{2009MNRAS.397.2099A}, respectively.
The MiNDSTEp data were reduced using DanDIA \citep{2008MNRAS.386L..77B, 2013MNRAS.428.2275B}.
$\mu$FUN data were reduced using DoPHOT \citep{1993PASP..105.1342S}, and LCO data from the LCO-\Spitzer\ program were reduced using a modified ISIS package \citep{1998ApJ...503..325A, 2000A&AS..144..363A, 2018PASP..130j4401Z}. 
The LCO data obtained by the ROME/REA team were reduced using a customized version of the DanDIA photometry pipeline. 
The $Spitzer$ data were reduced using the photometry algorithm described in \citet{2015ApJ...814...92C}.

It is known that the photometric error bars calculated by the data pipelines can be underestimated (or more rarely overestimated).
Various reasons, such as observational conditions, can cause systematic errors. In order to get proper errors of the parameters in the light-curve modeling, we empirically normalize the error bars by using the standard method of \citet{2008ApJ...684..663B}.
We use the formula,
\begin{equation}
  \sigma^{\prime}_{i} = k \sqrt{\sigma^{2}_{i} + e^{2}_{\rm min}},
\end{equation}
where $\sigma^{\prime}_{i}$ is the $i$th renormalized error, $\sigma_{i}$ is the $i$th error obtained from DIA, and $k$ and $e_{\rm min}$ are the renormalizing parameters. 
We set the value of $e_{\rm min}$ to account for systematic errors which dominate at high magnification, and we adjust the value of $k$ to achieve $\chi^2 {\rm /dof} = 1$.
The data from Salerno, LCO SAAO 
by the LCO-Spitzer program, LCO SSO and SAAO by the ROME/REA project are either too few to give any significant constraint or show systematics and disagreement with other datasets.
Therefore, we do not use them for the modeling.
We list the calculated error-bar renormalization parameters in Table \ref{Table_tel}.
\\

\begin{table}
\caption{The number of data points in the light curves and the normalization parameters}
\label{Table_tel}
\begin{center}
 \begin{tabular}{lllrrrrrr}
 \hline \hline
Name 			& Site 			& Collaboration	& Aperture(m) 	& Filter 	& $k$ & $e_{\rm min}$          &  $N_{\rm use} / N_{\rm obs}$    \\  \hline 
 OGLE   			& Chile			& OGLE 			&1.3  			& $ I$      &   1.660 & 0.003   &    3045/3045   \\  
 OGLE   			& Chile			& OGLE 			&1.3 			&  $V$      & 1.301 & 0.003   &     68/68   \\   
 MOA  			& New Zealand	& MOA 			& 1.8 			& $R_{\rm MOA}$         & 1.650  & 0.003   &  7277/7509               \\           
 MOA  			& New Zealand	& MOA 			& 1.8  			& $V$         &  1.321 & 0.003   &  240/240                \\       
 KMT SSO f03 	& Australia		&KMTNet 		& 1.6  			& $I$ &1.900 & 0.003 & 2087/2706 \\ 
 KMT SSO f43 	& Australia		&KMTNet 		& 1.6  			& $I$ & 1.824 & 0.003 &  2080/2658\\ 
 KMT CTIO f03 	& Chile 			&KMTNet		&1.6 			& $I$ & 1.579 & 0.003 & 2304/2486 \\ 
 KMT CTIO f43 	& Chile 			&KMTNet		&1.6 			& $I$ &1.443 & 0.003 &  2195/2363\\ 
 KMT SAAO f03 	& South Africa 	&KMTNet		&1.6 			& $I$ & 2.444 & 0.003 &  1813/2096\\ 
 KMT SAAO f43 	& South Africa 	&KMTNet		&1.6 			& $I$ & 1.900 & 0.003 &  1846/2078\\ 
 Danish 			& Chile			& MiNDSTEp		& 1.54 			& $Z$ & 1.015 & 0.003 &  139/154\\
 Salerno 			& Italy			& MiNDSTEp		& 0.6 			& $I$ & &  & 0/5\\
 LCO SSO 		& Australia		&LCO-$Spitzer$ 		& 1.0  			& $i^{\prime}$ & 2.528 & 0.003 &  31/44\\
 LCO CTIO		& Chile			&LCO-$Spitzer$ 		& 1.0  			& $i^{\prime}$ & 1.129 & 0.003 & 17/17 \\ 
 LCO SAAO 	& South Africa 	& LCO-$Spitzer$		&1.0 			& $i^{\prime}$ & & &  0/19\\
 CTIO 1.3m 		& Chile 			& $\mu$FUN 	&1.3 			& $I$ &  0.852 & 0.003 &  18/18\\ 
 CTIO 1.3m 		& Chile 			& $\mu$FUN 	&1.3 			& $V$ & 0.566 & 0.003 &  3/3\\ 
 LCO SSO 		& Australia		&ROME/REA		& 1.0 			& $g$ & &  &  0/25\\ 
 LCO SSO 		& Australia		&ROME/REA		& 1.0 			& $i^{\prime}$ & & &  0/74\\ 
 LCO SSO 		& Australia		&ROME/REA		&1.0 			& $r$ & & &  0/29\\ 
 LCO CTIO 	& Chile			&ROME/REA		& 1.0 			& $g$ &1.110 &0.003 &  33/33\\ 
 LCO CTIO 	& Chile			&ROME/REA		& 1.0 			& $i^{\prime}$ &1.589 &0.003 &  61/61\\ 
 LCO CTIO 	& Chile			&ROME/REA		& 1.0 			& $r$ &1.337 &0.003 &  31/31\\ 
 LCO SAAO	& South Africa	&ROME/REA		& 1.0 			& $g$ &  &  &  0/17\\ 
 LCO SAAO 	& South Africa	&ROME/REA		& 1.0 			& $i^{\prime}$ &  &  &  0/19\\ 
 LCO SAAO 	& South Africa	&ROME/REA		& 1.0			& $r$ &  &  &  0/45\\ 
 $Spitzer$		& Earth-trailing orbit & $Spitzer$ &0.85 			& $L$ &		2.110	&		\nodata &	26/26 \\
 \hline \hline
  \end{tabular}
  \end{center}
\end{table}

\section{Ground-based Light Curve Analysis}
\label{sec-analysis}

\subsection{Binary-lens model}
The magnification of the binary lens model depends on seven parameters: the time of lens-source closest approach $t_{0}$, the Einstein radius crossing time $t_{\rm E}$, the impact parameter in units of the Einstein radius $u_{0}$, the planet-host mass ratio, $q$, the planet-host separation in units of the Einstein radius, $s$, the angle between the trajectory of the source and the planet-host axis, $\alpha$, and the ratio of the angular source size to the angular Einstein radius, $\rho$.
The model flux $f(t)$ of the magnified source at time $t$ is given by,
\begin{equation}
f(t) = A(t) f_S +f_b,
\end{equation}
where $A(t)$ is a magnification of the source star, and $f_S$ and $f_b$ are the unmagnified flux from the source and the flux from any unresolved blend stars, respectively.

We also adopt a linear limb-darkening model for the source star,
\begin{equation}
S_\lambda(\vartheta) = S_\lambda(0)[1-u_\lambda(1-\cos (\vartheta))],
\end{equation}
where $S_\lambda(\vartheta)$ is a limb-darkened surface brightness.
The effective temperature of the source star estimated from the extinction-free source color presented in Section \ref{sec-cmd} is $T_{\rm eff} \sim 5662$K \citep{2009A&A...497..497G}. 
Assuming a surface gravity $\log g = 4.5$ and a metallicity of $\log [M/H] = 0$, we select the limb-darkening coefficients to be 
$u_I = 0.5494$,  $u_V = 0.7105$, $u_R = 0.6343$, $u_Z = 0.6314$, $u_g = 0.7573$, $u_r = 0.6283$ and $u_i = 0.5389$ from the ATLAS model \citep{2011A&A...529A..75C}. For the $R_{\rm MOA}$ passband, we use the coefficient for $u_{Red} = 0.5919$, which is the mean of $u_I$ and $u_R$.

We first conducted the light-curve fitting with only ground-based data.
We employed the Markov Chain Monte Carlo algorithm \citep{2003ApJS..148..195V} combined with the image-centered ray-shooting method \citep{1996ApJ...472..660B, 2010ApJ...716.1408B}. 
We conducted the grid search analysis following the same procedure in \citet{2019AJ....158..224K}. 
First, we performed a broad grid search over $(q, s, \alpha)$ space with the other parameters free. 
The search ranges of $q, s,$ and $\alpha$ are $-6 < \log q < 0$, $-0.5 < \log s < 0.6$, and $0 < \alpha < 2\pi$, with 11, 22, and 40 grid points, respectively. 
Next, we refined all parameters for the best 100 models with the smallest $\chi^{2}$ to search for the global best-fit model.

\begin{figure}[h]
\begin{center}
      \includegraphics[scale=0.6, angle=270]{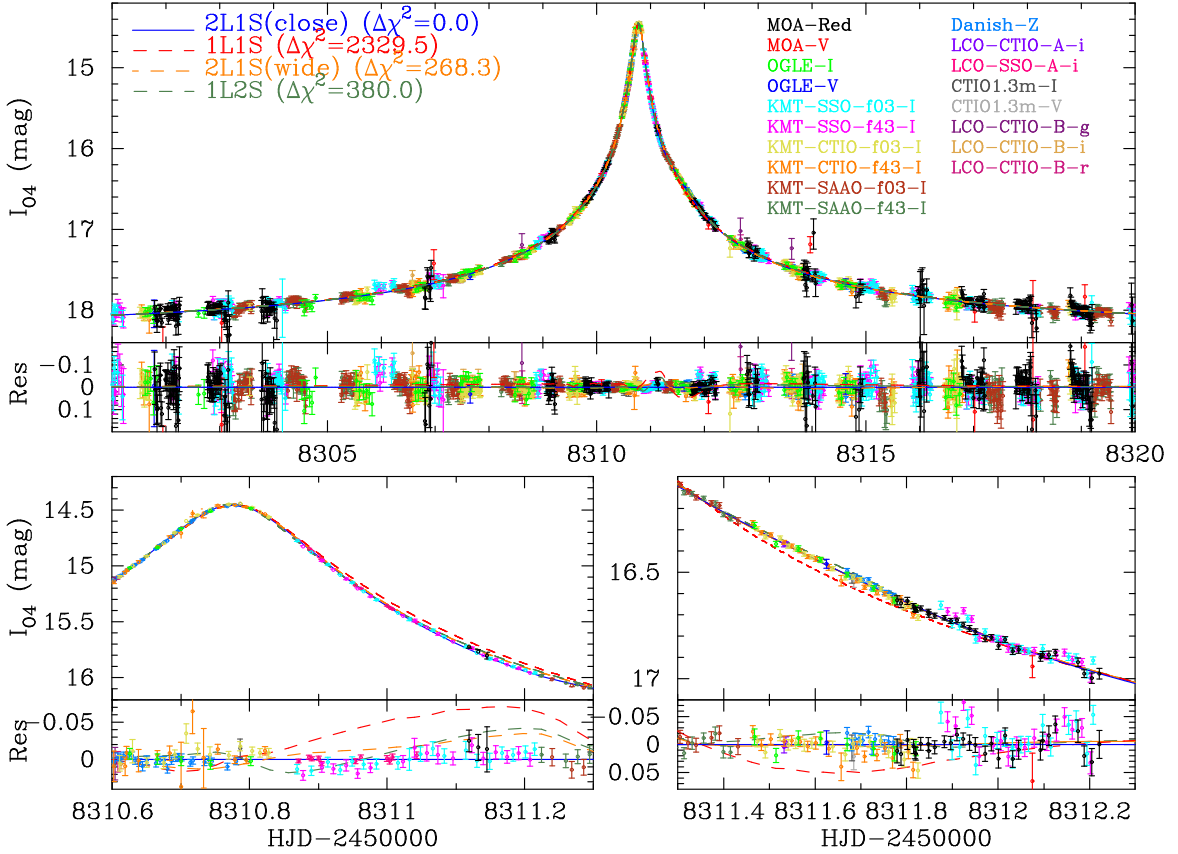}
   \caption{The light curve and models with the ground-based data for OGLE-2018-BLG-1185. Top panel shows the light curve, models, and residuals from the best-fit close binary-lens (2L1S) model. The blue line shows the best-fit close binary-lens (2L1S) model. The red, orange, and green dot lines show the single-lens (1L1S) model, the wide 2L1S model, and the binary-source (1L2S) model, respectively.
The left bottom panel and the right panel show the zoom-in of the light curve, where we can find clear deviations of data points from the 1L1S and 1L2S models.}
\end{center}
\label{Figure_lc}
\end{figure}

\begin{figure}[h]
\begin{center}
    \includegraphics[scale=0.5, angle=270]{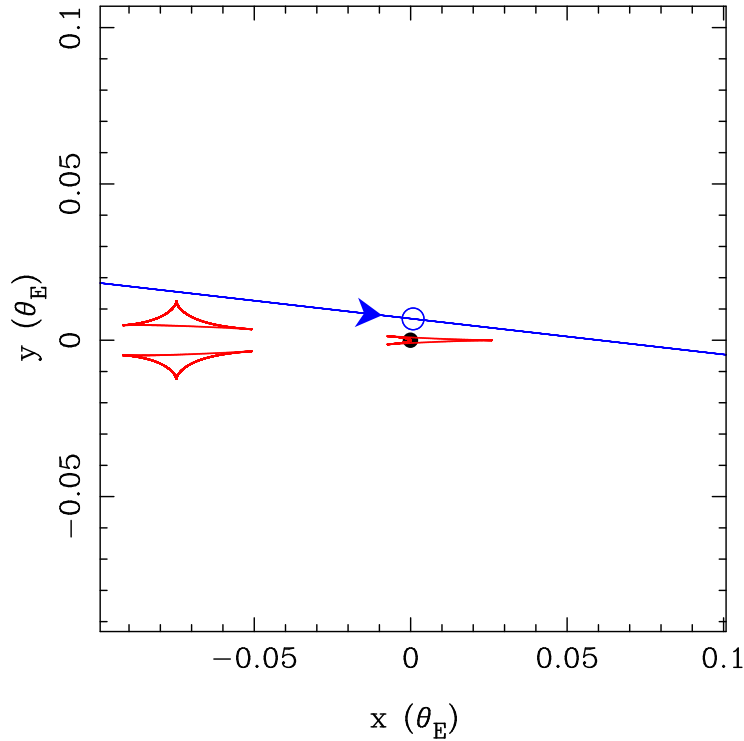}
   \caption{Caustic geometry of the best-fit model. The caustics are shown in red lines. The blue line shows the source trajectory on the lens plane and the arrow indicates the direction of the source/lens relative proper motion. The blue open circle indicates the source size and position at $t_0$.}
\end{center}
\label{Figure_cau}
\end{figure}

The parameters of the best-fit models are summarized in Table \ref{Table_best}. The light curve and the caustic geometry are shown in Figure \ref{Figure_lc} and Figure \ref{Figure_cau}.
As a result of the grid search, we found that the best-fit binary-lens model is favored over the single-lens model by $\Delta \chi^2 \sim 2330$. 
The bottom panels in Figure \ref{Figure_lc} show the clear deviations of the light curve with respect to the single-lens model from ${\rm HJD'} \sim 8310.9$ to $\sim 8311.8$, which are well fitted by the approach to the central caustic for the best binary-lens model.
Although the additional magnification from the cusp approach to the planetary caustic is small, the asymmetric feature in the right side of the light curve due to the approach to the central caustic shows clear residuals from the single-lens model, which suggests the existence of a companion.
The best binary-lens model suggests that the lens system has a very low mass ratio, $q \sim 6.9 \times 10^{-5}$, with a normalized separation $s \sim 0.96$. 
It is well known that there is a close/wide degeneracy in high-mag binary-lens events \citep{1998ApJ...500...37G, 1999A&A...349..108D, 2005ApJ...630..535C}, which is due to the similar shape and size of the central caustic between $s$ and $s^{-1}$. 
From the grid search, we found the best wide binary-lens model ($s>1$) has $q \sim 9.2 \times 10^{-5}$ and $s \sim 1.14$. The separation of this wide model is slightly different from the reciprocal of the separation of the close model ($s<1$), yielding a different shape and size for the central caustic from those of the best close model.
We ruled out the wide model because the best close binary-lens model is favored over the wide model by $\Delta \chi^2 \sim 268$.
The $\Delta \chi^2$ is large because the source trajectory is parallel to the lens axis and approaches not only the central caustic but also the planetary caustics.


\subsection{Binary-source model}
We checked the possibility that the observed light-curve can be explained by the binary-source (1L2S) model because it is known that there is a possible degeneracy between single-lens binary-source (1L2S) model and binary-lens single-source (2L1S) model \citep{1993ApJ...407..440G, 1998ApJ...506..533G}. 
For the 1L2S model, the total effective magnification of the source stars $A$ is expressed as follows, 
\begin{equation}
  A = \frac{A_{1}f_{1} + A_{2}f_{2}}{f_{1} + f_{2}} = \frac{A_{1} + q_{f}A_{2}}{1 + q_{f}},
\end{equation}
 where $A_{1}$ and $A_{2}$ are the magnification of the two sources with model flux $f_{1}$ and $f_{2}$, respectively, and $q_{f}$ is the flux ratio between the two sources ($=f_{2}/f_{1}$). 
In order to explain the magnification of the second source, we introduce the additional parameters: the time of lens-source closest approach $t_{0,2}$, the impact parameter in units of the Einstein radius $u_{0,2}$, and the ratio of the angular source size to the angular Einstein radius, $\rho_{2}$.
We found the best-fit 1L2S model is disfavored relative to the best-fit 2L1S model by $\Delta \chi^2 \sim380$, and we excluded the 1L2S model. The parameters of the best-fit 1S2L model are summarized in Table \ref{Table_best}. The light curve of the 1L2S model is shown in Figure \ref{Figure_lc}.

\subsection{Ground-Based Parallax}
The magnification of the binary lens model with parallax effects need two additional parameters: the North and East components of parallax vector $\bpi_{\rm E}$ in equatorial coordinates, $\pi_{\rm E,E}$ and $\pi_{\rm E,N}$ \citep{2004ApJ...606..319G}.
The orbital parallax effects are caused by Earth's orbital motion. 
In the case of OGLE-2018-BLG-1185, the timescale, $t_{\rm E}\sim 15.9$ days, is small compared to Earth's orbital period, which makes it less likely to measure the parallax effects. 
The best-fit parallax model improves the fit slightly by $\Delta \chi^2\sim 20$, but there is disagreement of the $\chi^2$ improvement between the datasets.
The parallax information such as the direction and the value is easily influenced by the systematics in each telescope dataset.
Considering these facts, we concluded that we should disregard the parallax information from the ground-based data.

\begin{center}
\begin{threeparttable}
\caption{The best-fit models for ground-only data}
\label{Table_best}
\begin{tabular}{cc|rrr}
 \hline \hline
 Parameters				        &Unit			    &  2L1S(close) 					  & 2L1S(wide) 				    &1L2S                                   \\  \hline 
 $\chi^2/{\rm dof}$		    & 				    & 23221.473/23252 				&23489.306/23252			  &23601.431/23249 \\
$t_{0,1}$ 				        &HJD$^\prime$	& 8310.7772 $\pm$ 0.0003 	&8310.7793 $\pm$ 0.0003	&8310.7726 $\pm$  0.0003\\       
$t_{0,2}$ 				        &HJD$^\prime$	& ... 								    & 	...						      &8311.5874 $\pm$  0.0010\\     
$t_{\rm E}$ 		  		    &days			    & 15.931 $\pm$ 0.133			&16.312 $\pm$ 0.144			&15.730 $\pm$  0.189	                 \\          
$u_{0,1}$ 			          &$10^{-3}$		& 6.877 $\pm$ 0.063				&6.606 $\pm$ 0.067			&7.777 $\pm$  0.131  	      \\         
$u_{0,2}$ 			          &$10^{-3}$		& 	...							      &	...						        &8.773 $\pm$ 1.515					     \\   
$q$ 				              &$10^{-5}$		& 6.869 $\pm$ 0.229 			&9.164 $\pm$ 0.552			& ... 	       \\         
$s$              			    &				      & 0.963 $\pm$ 0.001 			&1.144 $\pm$ 0.003			&  ...        \\         
$\alpha$ 		        	    &radian		   	& 0.114 $\pm$ 0.001 			&3.261 $\pm$ 0.002			&  ...        \\    
$\rho_{1}$ 			          &$10^{-3}$		& 3.468 $\pm$ 0.083				& $<$1.026\tnote{a}  	  &7.234 $\pm$  0.241	\\      
$\rho_{2}$ 			          &$10^{-3}$		& 	...							      &	...						        &1.613 $\pm$  0.956  	       \\    
$q_{f,I}$					        &$10^{-2}$		&	...							        &	...						        &1.699 $\pm$  0.192  	\\
$f_S$ (OGLE)	\tnote{b}		&				      & 107.777 $\pm$ 0.437			&106.493 $\pm$ 0.448		&108.583 $\pm$ 0.550	  \\
$f_b$ (OGLE)	\tnote{b}		&				      & 396.165 $\pm$ 0.594			&397.397 $\pm$ 0.440		&393.516 $\pm$ 0.587	  \\
 \hline \hline
\end{tabular}
\begin{tablenotes}
\item[a] The value is the $3\sigma$ upper limit.
\item[b] All fluxes are on a 25th magnitude scale, e.g., $I_{\rm S}=25-2.5\log(f_{\rm S})$.
\end{tablenotes}
\end{threeparttable}
\end{center}

\section{Angular Einstein Radius}
\label{sec-cmd}

We can estimate the angular Einstein radius $\theta_{\rm E}=\theta_{\rm *}/\rho$ because $\rho$ can be derived by the light-curve fitting and the angular source radius $\theta_\ast$ can be derived by using an empirical relation between $\theta_\ast$, the extinction-corrected source color $(V-I)_{\rm S,0}$, and the magnitude $I_{\rm S,0}$ \citep[e.g., ][]{2014AJ....147...47B}.

We derived the OGLE-IV instrumental source color and magnitude from the light-curve fitting and then converted them to the standard ones by using the following color-color relation 
from \citet{2015AcA....65....1U}:
\begin{eqnarray}
I_{\rm O3} - I_{\rm O4} &=& (0.182 \pm 0.015) + (-0.008 \pm 0.003)(V - I)_{\rm O3}, \\
V_{\rm O3} - V_{\rm O4} &=& (0.257 \pm 0.015) + (-0.074 \pm 0.004)(V - I)_{\rm O3}.
\end{eqnarray}
The apparent color and the standard magnitude of the source star are $(V-I,I)_{\rm S, O4, calib} = (2.344 \pm 0.031, 20.082 \pm 0.012)$.

We also derived the apparent source color and magnitude from the CT13 measurements in the $I$- and $V$-bands from the light-curve fitting, and then converted them to the standard ones following the procedure explained in  \citet{2017MNRAS.469.2434B}. We cross-referenced isolated stars in the CT13 catalog reduced by DoPHOT \citep{1993PASP..105.1342S} with the stars in the OGLE-III map within $120^{\prime\prime}$ of the source star and obtained the color-color relation:
\begin{eqnarray} 
I_{\rm O3} - I_{\rm CT13} &= &(-0.880 \pm 0.005) + (-0.042 \pm 0.005)(V - I)_{\rm CT13}, \\
V_{\rm O3} - I_{\rm CT13} &=& (1.290 \pm 0.004) + (-0.036 \pm 0.004)(V - I)_{\rm CT13}.
\end{eqnarray}
The apparent color and magnitude of the source star are $(V-I,I)_{\rm S, CT13, calib} = (2.335 \pm 0.025, 20.105 \pm 0.013)$.
This color is consistent with $(V-I)_{\rm S, O4, calib}$ within $1\sigma$ and the magnitude is consistent with $I_{\rm S, O4, calib}$ within $2\sigma$. Because the light curve was well covered by the OGLE observations, while it was highly magnified, we adopted $(V-I,I)_{\rm S, O4, calib}$
as the source color and magnitude.

\begin{figure}[h]
\begin{center}
    \includegraphics[scale=0.7, angle=270]{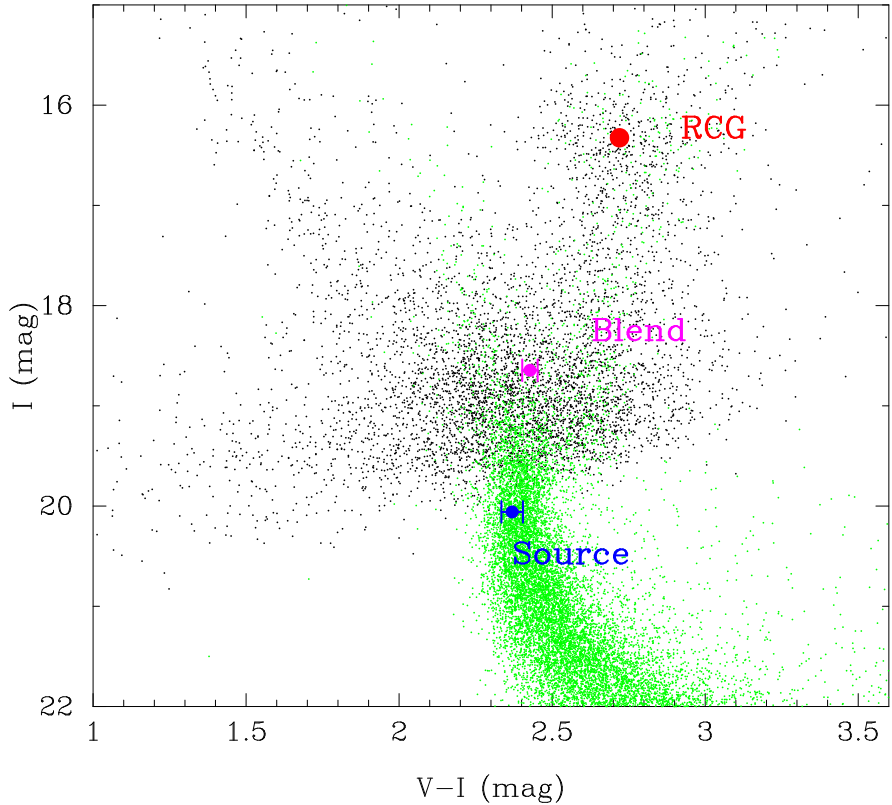}
   \caption{Color magnitude diagram (CMD). The stars in the OGLE-III catalog within $120^{\prime\prime}$ of the source star are shown with black dots. The green dots indicate the $HST$ CMD of \citep{1998AJ....115.1946H}, which is transformed to the same reddening and extinction of the field of the event. The red dot shows the centroid of the red clump giant distribution. The colors and magnitudes of the source star and the blend are shown with blue and pink, respectively.}
\end{center}
\label{Figure_cmd}
\end{figure}

To obtain the extinction-corrected source color and magnitude, we used the standard method from \citet{2004ApJ...603..139Y}. The intrinsic color and magnitude are determined from the source location relative to the color and magnitude of the red clump giant (RCG) centroid in the color-magnitude diagram (CMD). In Figure \ref{Figure_cmd}, the red point shows the RCG centroid color and magnitude, $(V-I,I)_{\rm RCG} = (2.720, 16.325) \pm (0.009, 0.032)$ for the field around the source star. 
Assuming that the source star suffers the same reddening and extinction as the RCGs, we compared these values to the expected extinction-corrected RCG color and magnitude for this field, $(V-I,I)_{\rm RCG,0} = (1.060, 14.362) \pm (0.070, 0.040)$ \citep{2013A&A...549A.147B, 2013ApJ...769...88N}. As a result, we obtained an extinction of $A_I = 1.963 \pm 0.051$ and a color excess of $E(V-I) = 1.660 \pm 0.071$.
Finally, the intrinsic source color and magnitude were derived,
\begin{equation}
(V-I, I)_{{\rm S},0} = (0.684, 18.119) \pm (0.077, 0.053).
\end{equation}
As a reference for the later discussion of the future follow-up observations, we also estimated the intrinsic source magnitudes in $H$- and $K$-bands from the color-color relation in \citet{1995ApJS..101..117K}, including a 5\% uncertainty. Then, we applied the extinction in the $H$- and $K$-bands, which were derived from the extinction in the $I$- and $V$-bands of the RCGs according to \citet{1989ApJ...345..245C}.

Figure \ref{Figure_cmd} shows that the source is consistent with being part of the standard bulge sequence of stars, i.e., it falls within the distribution of stars from \citep{1998AJ....115.1946H} after they have been shifted to the same reddening and extinction as the field for \thisevent. However, the source also has a similar color to the Sun. Thus, it would also be consistent with being a similar absolute magnitude to the Sun but  somewhat in the foreground, e.g., at $\sim 6$ kpc. Thus, we also checked how a different assumption about the source would affect our results. If the source was more in the foreground, it would then suffer less extinction and reddening than the RCGs. 
However, even if we assume 10\% less extinction and reddening than the RCGs, the value of $\theta_{\rm E}$ increases by only 7\%,  which is still consistent within $1\sigma$ with values obtained assuming the same extinction and reddening as the RCGs.
We summarize the source color and magnitudes in Table \ref{Table_cmd}.

Applying the empirical formula, $\log(\theta_{\rm LD}) = 0.501414 + 0.419685(V-I) - 0.2I$ (see \citealp{2015ApJ...809...74F} but also \citealp{2014AJ....147...47B}), 
where $\theta_{\rm LD} \equiv 2\theta_*$ is the limb-darkened stellar angular diameter, we found the angular source radius,
\begin{eqnarray} 
 {\bf \theta_{\rm LD}} &=& 1.461 \pm 0.109 \ {\rm \mu as},\\
  \theta_{\rm *} &=& 0.730 \pm 0.059 \ {\rm \mu as}.
 \end{eqnarray}
Finally, we obtained the source angular radius and the lens–source relative proper motion in the geocentric frame,
\begin{eqnarray} 
\theta_{\rm E} &=& \frac{\theta_*}{\rho} = 0.211\pm 0.018\ {\rm mas}, \\
\mu_{\rm rel, geo}  &=& \frac{\theta_{\rm E}}{t_{\rm E}} = 4.832 \pm 0.410\ \mathrm{mas\ yr}^{-1}
\end{eqnarray}
This $\theta_{\rm E}$ value is relatively small, which suggests that the lens is a low-mass star and/or distant from the observer.
\\

\begin{center}
\begin{threeparttable}
\caption{The source color and magnitudes}
\label{Table_cmd}
\begin{tabular}{cc|ccc}
 \hline \hline
 Parameters				&Unit&      Source &    Source &    Source                          \\ 
 			&&     (apparent) &   (intrinsic)\tnote{a}&   (intrinsic)\tnote{b}                          \\ \hline 
 $I$						&mag					&20.082	$\pm$ 0.012 \tnote{c}	&18.119$\pm$0.053		&18.315$\pm$0.053	 \\
$V-I$ 				   	&mag					&2.344 $\pm$ 0.031 \tnote{c}   &0.684$\pm$0.077		&0.850$\pm$0.077           \\        
\hline
$H$	 \tnote{d} 			&mag					&18.012$\pm$0.143  &17.444$\pm$0.095		& \nodata \\
$K$	 \tnote{d}			&mag					&17.756$\pm$0.145   &17.394$\pm$0.095		& \nodata \\
 \hline \hline
\end{tabular}
\begin{tablenotes}
\item[a] Extinction-corrected magnitudes assuming that the source star suffers the same reddening and extinction as the RCGs.
\item[b] Extinction-corrected magnitudes assuming that the source star suffers the reddening and extinction of 0.9 times as much as that for the RCGs.
\item[c] The magnitude and color are measured from the light-curve fitting.
\item[d] The magnitudes are estimated from the color-color relation in \citet{1995ApJS..101..117K} and the extinction law in \citet{1989ApJ...345..245C}.
\end{tablenotes}
\end{threeparttable}
\end{center}

\section{Lens Physical Parameters by Bayesian Analysis}
\label{sec-bay}
If we can measure both the finite source effects and the parallax effects, 
the lens physical parameters such as the host mass $M_{\rm host}$ and the distance to the lens $D_{\rm L}$ are calculated directly, following the equations:
\begin{equation}
M_{\rm host}= \frac{\theta_{\rm E}}{(1+q)\kappa\pi_{\rm E}};\ \ D_{\rm L} = \frac{\rm au}{\pi_{\rm rel}+\pi_{\rm S}};\ \ \pi_{\rm rel}= \theta_{\rm E}\pi_{\rm E};\ \ {\bm \mu}_{\rm rel} = \frac{\theta_{\rm E}}{t_{\rm E}}\frac{{\bm \pi}_{\rm E}}{\pi_{\rm E}},
\label{eqn:ML}
\end{equation}
where $\kappa \equiv 4G/(c^2{\rm au})=8.1439\ {\rm mas}/M_\odot$, and $\pi_{\rm S}={\rm au}/D_{\rm S}$ is the source parallax.
From the ground-based light-curve alone, we are only able to measure $\theta_{\rm E}$ (via finite source effects), but no meaningful constraint on $\pi_{\rm E}$ (see Section 3.3).

In order to estimate the probability distributions of $M_{\rm L}$ and $D_{\rm L}$, we conducted a Bayesian analysis with the Galactic model of \citet{2021arXiv210403306K} as a prior. We randomly generated 50 million simulated microlensing event samples. Then we calculated the probability distributions for the lens physical parameters by weighting the microlensing event rate by the measured $t_{\rm E}$ and $\theta_{\rm E}$ likelihood distribution. 
It is important to note that we conducted the Bayesian analysis under the assumption that stars of all masses have an equal probability to host this planet. 

We calculated some parameters in addition to the lens physical parameters, $M_{\rm L}$ and $D_{\rm L}$.
For instance, the lens-source proper motion in the geocentric frame, ${\bm \mu}_{\rm rel}$, is converted to that in the heliocentric frame,
\begin{equation}
{\bm \mu}_{\rm rel,hel} = {\bm \mu}_{\rm rel} + {\bm v}_{\oplus,\perp} \frac{\pi_{\rm rel}}{\rm au},
\label{eqn:muhel}
\end{equation}
where ${\bm v}_{\oplus,\perp} = ({v}_{\oplus,N},{v}_{\oplus, E}) = (-0.78, 27.66)$ km/s is the projected velocity of Earth at $t_0$. 

We also calculated the $I$- and $V$-band magnitudes of the lens from the mass-luminosity relations of main-sequence stars \citep{1995ApJS..101..117K}, and the 5 Gyr isochrone for brown dwarfs from \citet{2003A&A...402..701B}. Then we estimated $H$- and $K$-band magnitudes of the lens from the color-color relation in \citet{1995ApJS..101..117K}, including a 5\% uncertainty. In order to estimate the extinction in the foreground of the lens, we assumed a dust scale height of $h_{\rm dust} = 0.10 \pm 0.02$ kpc \citep{2015ApJ...808..169B},
\begin{equation}
A_{\lambda, {\rm L}} = \frac{1-e^{-|D_{\rm L}/(h_{\rm dust}\sin{b})|}}{1-e^{-|D_{\rm S}/(h_{\rm dust}\sin{b})|}} A_{\lambda,\rm{S}},
\end{equation}
where the index $\lambda$ refers to the passband: $V$-, $I$-, $H$-, or $K$-band. We obtained the extinction in the $I$- and $V$-band magnitudes of the source from the RCGs in Section \ref{sec-cmd}, and then we converted them to the extinction in the $H$- and $K$-bands according to \citet{1989ApJ...345..245C}.

\begin{figure}[h]
\begin{center}
    \includegraphics[scale=0.55,angle=270]{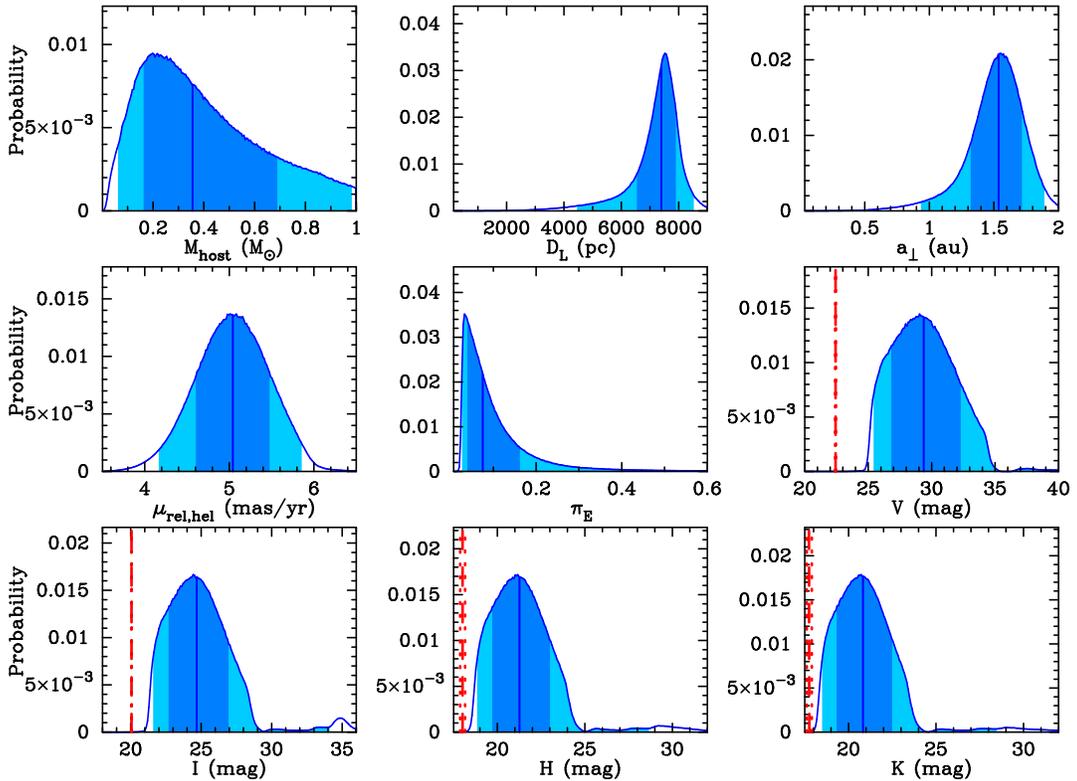}
\caption{Probability distribution of lens properties derived from the Bayesian analysis with a Galactic prior and constrained by $t_{\rm E}$ and $\theta_{\rm E}$.
The vertical blue lines show the median values. The dark-blue and the light-blue regions show the 68.3\% and 95.4\% confidence intervals. The red vertical lines in the probability distributions of $I$-, $V$-, $H$-, and $K$-band magnitudes show the magnitudes of the source star with extinction.}
\end{center}
\label{Figure_bay_gr}
 \end{figure}

\begin{figure}[h]
\begin{center}
    \includegraphics[height=0.4\textheight]{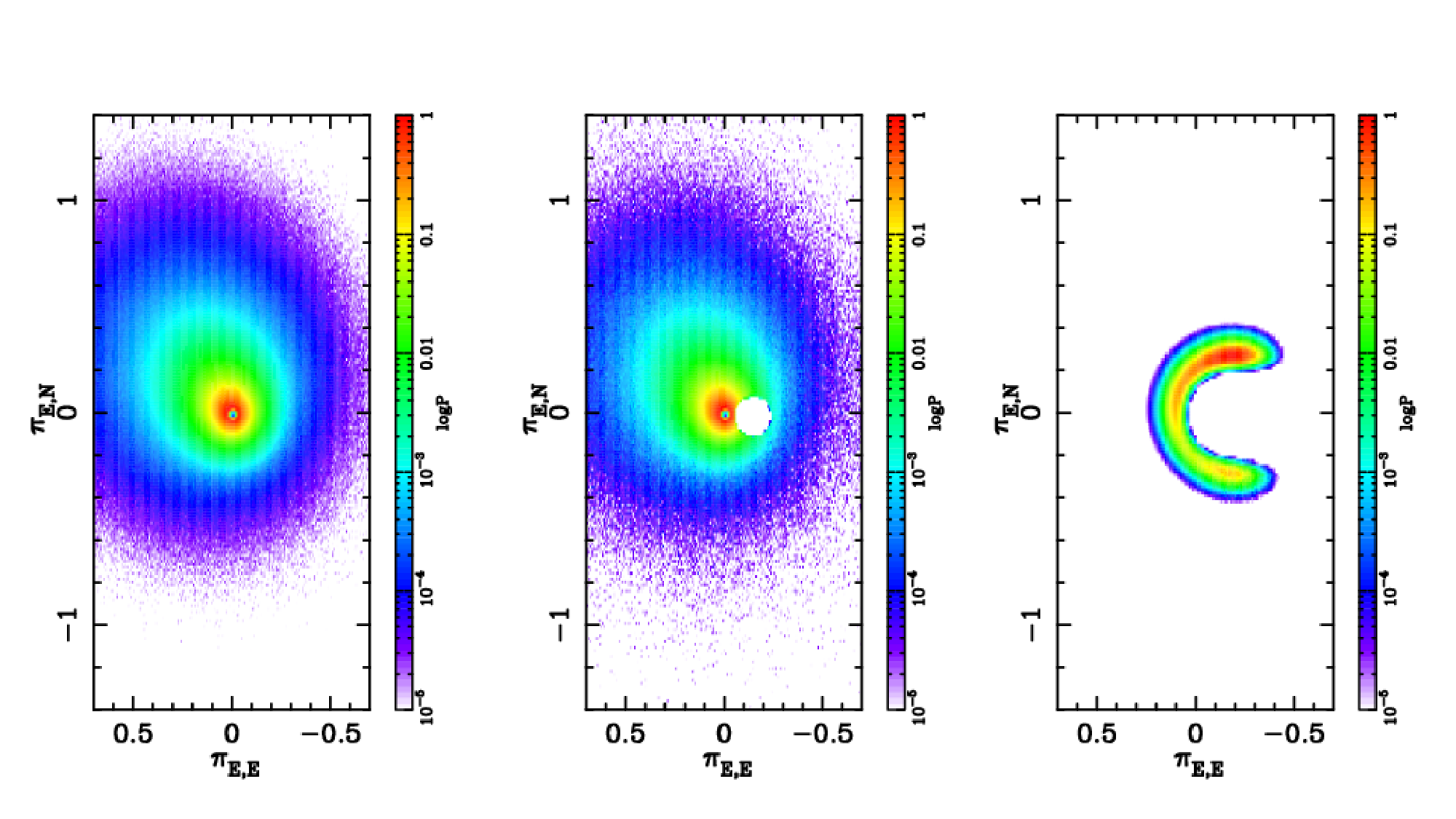}
    \caption{{\em Left}: The parallax contours for OGLE-2018-BLG-1185 expected from the Galactic model of \citet{2021arXiv210403306K} after imposing the two observational constraints of the angular Einstein radius, $\theta_{\rm E}$ and the Einstein radius crossing time, $t_{\rm E}$, on the event rate. The colorbar corresponds to the logarithm of the event rate and the red region indicates higher probability. {\em Center}: Including the constraint that $\Delta f_{\rm Spz} < 4$. {\em Right}: Including the full constraint from the \Spitzer-``only" parallax.}
\end{center}
\label{Figure_kos}
 \end{figure}

The results are shown in Table \ref{Table_bay} and Figure \ref{Figure_bay_gr}. 
According to Figure \ref{Figure_bay_gr}, the lens system is likely to be a super-Earth with a mass of $m_{\rm p} =8.1^{+7.6}_{-4.4}\ M_\oplus$ orbiting a late M-dwarf with a mass of $M_{\rm host} = 0.36^{+0.33}_{-0.19} M_\odot$ at a projected separation of $a_\perp = 1.54^{+0.18}_{-0.22}\ {\rm au}$. The system is
located $D_{\rm L} = 7.4^{+0.5}_{-0.9}$ kpc from Earth. 
For reference, we also plot the source magnitudes in $V$-, $I$-, $H$-, and $K$-band as the red lines; the $H$- and $K$-band magnitudes were estimated in Section \ref{sec-cmd}. 
We also show the parallax contour derived from the Bayesian analysis in Figure \ref{Figure_kos}.

 \begin{table}[h]
\caption{The lens physical parameters}
\begin{center}
\label{tab-para}
 \begin{tabular}{cc|rrr|rrrrrr}
 \hline \hline
   	   	                & &  \multicolumn{3}{c|}{Bayesian}	& 	  \multicolumn{2}{c}{Naive \Spitzer-``only"}\\ 
 Parameters 	   	    & Unit        		& Ground-only 	&  Ground + $\Delta f_{\rm Spz}$ & Ground + $\bpi_{\rm E, Spz}$ & $(u_0 > 0)$ & $(u_0 < 0)$\\  
 \hline
$M_{\rm host}$      	& $M_\sun$			&  
  	 $0.36^{+0.33}_{-0.19}$	 & $0.37^{+0.35}_{-0.21}$  	& $0.091^{+0.064}_{-0.018}$  &  $0.073 \pm 0.011$ & $0.070 \pm 0.010$\\         
$m_{\rm p}$   				& $M_\oplus$		&  
  	 $8.1^{+7.6}_{-4.4}$ 	 & $8.4^{+7.9}_{-4.7}$ 	& $2.1^{+1.5}_{-0.4}$ &  $1.7 \pm 0.3$ & $1.6 \pm 0.2$\\             
$D_L$                 & kpc				&  
  	 $7.40^{+0.51}_{-0.85}$ 	 & $7.40^{+0.51}_{-0.88}$ 	& $5.45^{+1.70}_{-0.66}$ & $4.96 \pm 0.74$ & $4.89 \pm 0.66$\\  
$a_\perp$ 			& au				& 
  	 $1.54^{+0.18}_{-0.22}$ 	 & $1.54^{+0.18}_{-0.22}$ 	& $1.14^{+0.32}_{-0.15}$ & $1.01 \pm 0.18$ & $0.99 \pm 0.16$\\ 
$\pi_{\rm E}$			& 					&
 	 $0.075^{+0.087}_{-0.036}$ 	 & $0.073^{+0.093}_{-0.035}$  	& $0.292^{+0.066}_{-0.120}$ & $0.354 \pm 0.042$  & $0.369 \pm 0.037$\\
$\mu_{{\rm rel, hel}}$  & mas/${\rm yr}$	& 
	 $5.04^{+0.43}_{-0.44}$ 	 & $5.06^{+0.43}_{-0.44}$	& $4.86\pm0.44$ & \nodata & \nodata \\ 
$V$						& mag 				& 
	 $29.4^{+2.9}_{-2.6}$  & $29.3^{+3.1}_{-2.6}$	& $34.1^{+5.2}_{-1.6}$ & \nodata & \nodata \\
$I$						& mag 				&
	 $24.7^{+2.3}_{-2.0}$  & $24.6^{+2.4}_{-2.0}$	& $28.2^{+3.4}_{-1.2}$ & \nodata & \nodata \\
$H$						& mag 				&
	 $21.3^{+1.7}_{-1.6}$  & $21.2^{+1.9}_{-1.6}$	& $23.9^{+2.6}_{-0.9}$ & \nodata & \nodata \\
$K$						& mag 				&
	 $20.8^{+1.7}_{-1.5}$  & $20.8^{+1.8}_{-1.5}$	& $23.3^{+2.9}_{-0.8}$ & \nodata & \nodata \\
 \hline \hline
\end{tabular}
\label{Table_bay}
\end{center}
\end{table}

\newpage
\section{Analysis including \Spitzer\ data}
\label{sec-sp}

\begin{figure}[h]
\begin{center}
    \includegraphics[width=0.75\textwidth]{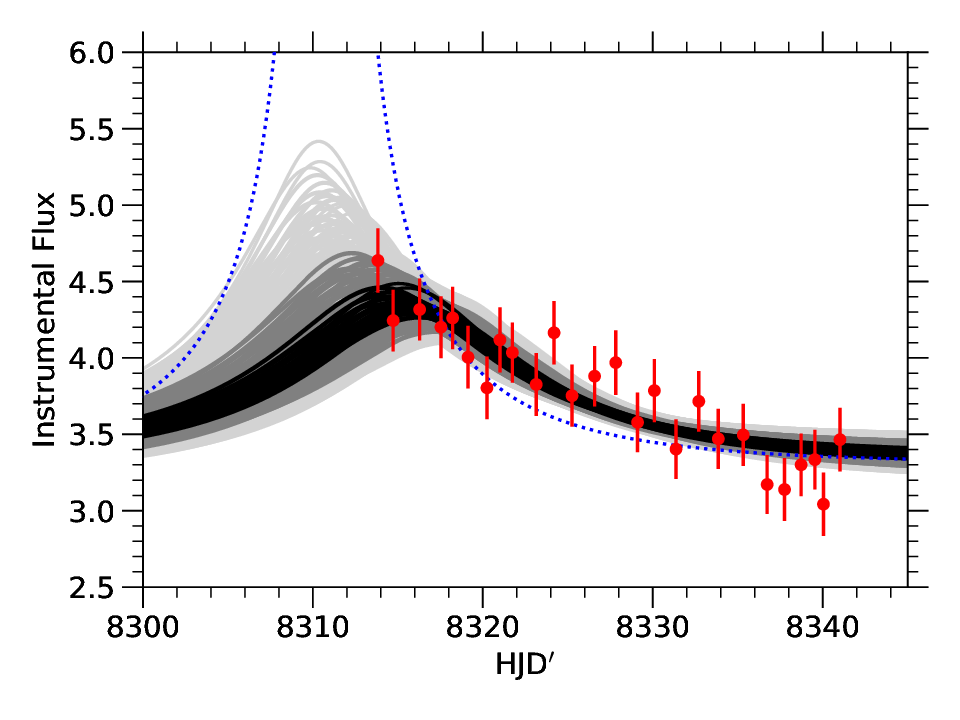}
   \caption{The light curve and models with the $Spitzer$ data. The blue dotted line shows the \Spitzer\ flux predicted by the 2L1S best-fit model derived from the ground-based analysis for $\bpi_{\rm E} = (0, 0)$ evaluated at the central value of the color-constraint. The black and gray shaded regions show models derived from the $Spitzer$-``only" parallax analysis. Each color (black, dark gray, light gray) represents $\Delta \chi^2 <$ (1,4,9).}
\end{center}
\label{Figure_splc}
\end{figure}

\begin{figure}[h]
\begin{center}
   \includegraphics[scale=0.55,angle=270]{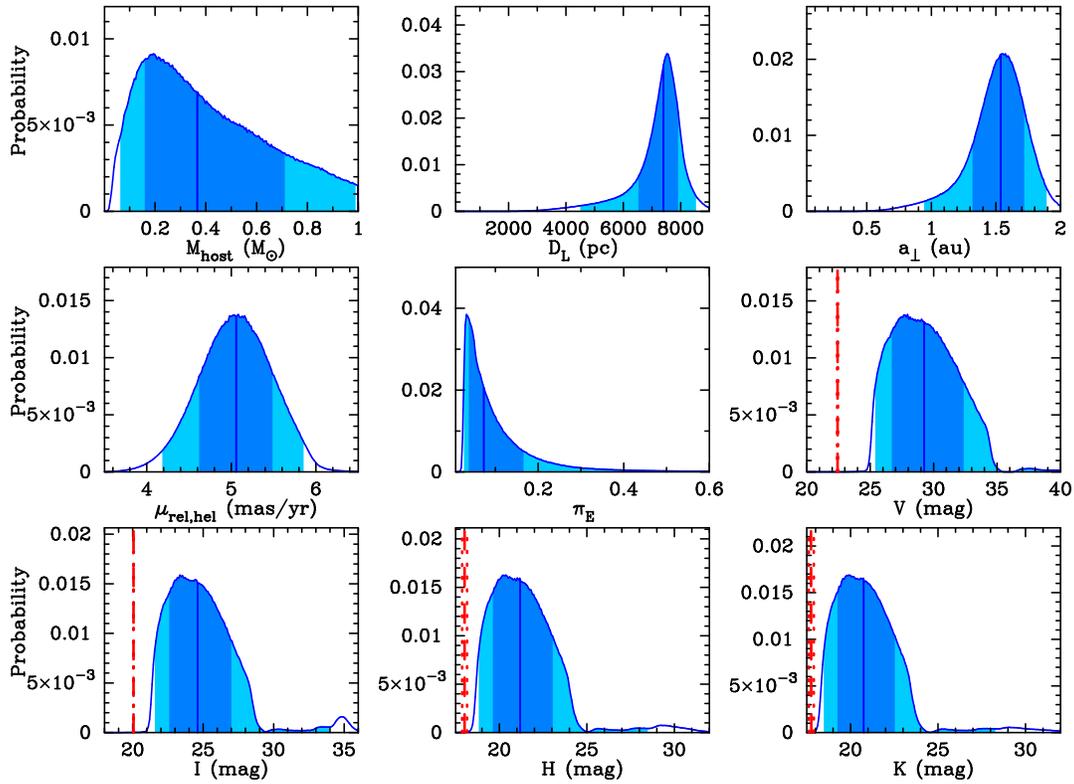}
	\caption{Same as Figure \ref{Figure_bay_gr}, but with the addition of the constraint $\Delta f_{\rm Spz} < 4$.}
\end{center}
\label{Figure_bayes_dflux}
 \end{figure}

We measure the microlens parallax vector $\bpi_{\rm E}$ via the ``satellite parallax effect'', which can be approximated as:
\begin{equation}
\bpi_{\rm E} = \frac{\rm au}{D_{\perp}}\left( \frac{t_{0,{\rm sat}}-t_{0,\oplus}}{t_{\rm E}}, u_{0,{\rm sat}}-u_{0,\oplus} \right),
\label{eqn:sat}
\end{equation}
where $D_{\perp}$ is the Earth-satellite separation projected on the plane of the sky, and $t_{0,{\rm sat}}$ and $u_{0,{\rm sat}}$ are the time of lens-source closest approach and the impact parameter as seen by the satellite. The Einstein timescale $t_{\rm E}$ is assumed to be the same for both Earth and the satellite. In practice, we fully model \Spitzer’s location as a function of time.

The \Spitzer\ light curve for \thisevent\ shows a very weak decline of $\Delta f_{\rm Spz} \sim 1$ flux unit over the four-week observation period (see Figure \ref{Figure_splc}). This change (rather than, e.g., the value of the flux at the start of observations) is the most robust constraint because it is independent of the unknown blended light. However, the magnitude of the decline is comparable to the level of systematics seen in a few other events \citep{2020JKAS...53....9G,2020AJ....160...74H,2020arXiv201008732Z} and, thus, should be treated with caution. At the same time, even this weak decline indicates a significant parallax effect for the event as seen from \Spitzer. We derive a color  constraint for the \Spitzer\ data by measuring the $IHL$ color-color relation for clump stars in CT13 $I$ and $H$, and \Spitzer\ $L$. Evaluating this relation at the measured $(I-H)$ color of the source gives the constraint on the \Spitzer\ source flux:
\begin{equation}
I_{\rm CT13} - L = -4.518 \pm 0.028,
\label{eqn:cc}
\end{equation}
which gives an expected source flux from \Spitzer\ of $f_{\rm S, Spz} = 0.6254$ flux units for the best-fit value of $I_{\rm CT13}$. This constraint and the best-fit ground-based model (Table \ref{tab-para}) together imply some tension with the observed \Spitzer\ light curve unless there is a significant parallax effect. They predict that the observed \Spitzer\ flux should have been substantially brighter at the start of the \Spitzer\ observations ($f_{\rm Spz}(\mathrm{HJD}^{\prime}=8313.83) \sim 6$ flux units) and declined by a total of $\Delta f_{\rm Spz} \sim 3.3$ flux units as compared to the observed $\Delta f_{\rm Spz} \sim 1$ flux unit. This tension can be seen in Figure \ref{Figure_splc} and suggests that, due to the parallax effect, the event peaked at a lower magnification and/or earlier as seen from \Spitzer.



We can use limits on the change in the \Spitzer\ flux ($\Delta f_{\rm Spz}$) to place conservative constraints on the physical properties of the lens.  Suppose that systematics affect the \Spitzer\ light curve at the level of 1--2 flux units, i.e., at the level seen in previous work. If the true signal is $\Delta f_{\rm Spz} \sim 4$ flux units, it is very unlikely that systematics would cause us to measure $\Delta f_{\rm Spz} = 1$ flux unit. Therefore, we repeat the Bayesian analysis imposing the constraint $\Delta f_{\rm Spz} < 4$, where $\Delta f_{\rm Spz}$ is calculated from Equation (\ref{eqn:cc}). The parallax effect can produce a degeneracy in the sign of $u_0$. In this case, because $u_0$ is small, the effect of this degeneracy is much smaller than the uncertainties \citep{GouldYee12}, so we only carry out this calculation for the $u_0 > 0$ case.

The results are given in Table \ref{Table_bay} (as ``Ground + $\Delta f_{\rm Spz}$"), Figure \ref{Figure_bayes_dflux}, and the center panel of Figure \ref{Figure_kos}. This constraint suggests a $M_{\rm host} = 0.37^{+0.35}_{-0.21}\ M_\odot$ host with a $m_p =  8.4^{+7.9}_{-4.7}\ M_\oplus$ planet at a projected separation $a_\perp = 1.54^{+0.18}_{-0.22}\ {\rm au}$. We adopt these values as our conservative Bayesian estimate of the properties of the lens system.

\subsection{Spitzer-``only" Parallax}

\begin{figure}
	\begin{centering}
	\includegraphics[height=0.5\textheight]{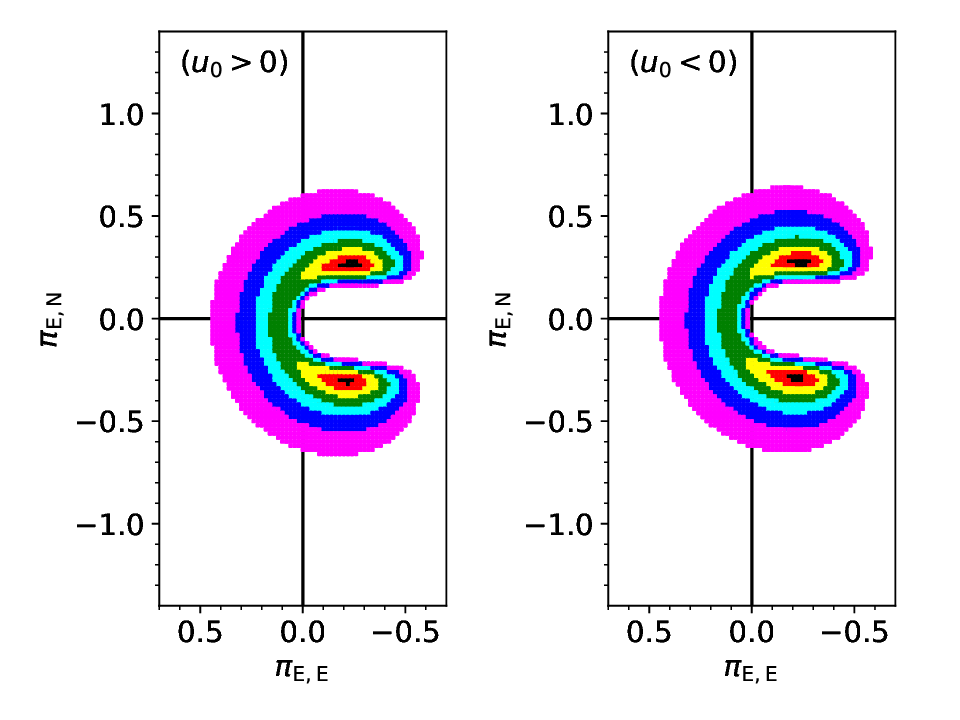}
	\caption{Parallax contours from \Spitzer-``only" analysis (see text). Colors (black, red, yellow, green, cyan, blue, magenta) indicate (1, 2, 3, 4, 5, 6, 7)$\sigma$ from the minimum, respectively. Left panel is for $(u_0 > 0)$ and right panel is for $(u_0 < 0)$. \label{fig:sponly}}
	\end{centering}
\end{figure}

If we take the \Spitzer\ light curve at face value, we can derive stronger constraints on the parallax using the \Spitzer-``only" parallax method. This method has been used in several previous analyses \citep[starting with][]{2020JKAS...53....9G} to show how the \Spitzer\ light curve constrains the parallax. For this analysis, we hold the microlensing parameters $t_0$, $u_0$, and $t_{\rm E}$ fixed at values found by fitting the ground-based data and make the assumption that the \Spitzer\ light curve is in the point lens regime.\footnote{In principle, we should calculate the \Spitzer\ magnification using the full planetary model, but in practice, this makes almost no difference because the \Spitzer\ observations start well after the planetary perturbation.} Then, for a grid of parallax values, we fit for the \Spitzer\ flux while applying the color-constraint from Equation (\ref{eqn:cc}).
 We repeat the analysis for $-u_0$, which produces an indistinguishable ground-based light curve and, as expected, only slight variations in the parallax.

The resulting parallax contours are shown in Figure \ref{fig:sponly}. The four minima correspond to the well-known satellite parallax degeneracy \citep{1966MNRAS.134..315R,Gould94} and the overall arc shape follows the expectation from the \citet{2019JKAS...52..121G} osculating circles formalism. The values for the magnitude of the microlens parallax vector are $\pi_{\rm E} = 0.35 \pm 0.04 $ for the $(u_0 > 0)$ case and $\pi_{\rm E} = 0.37 \pm 0.04$ for the $(u_0 < 0)$ case. The $3\sigma$ ranges are $\pi_{\rm E} = [0.18, 0.50]$ and $\pi_{\rm E} = [0.20, 0.48]$, respectively.
\\

{\subsection{Physical Lens Properties from Spitzer Parallax}
\label{sec:phys_sponly}}

We can derive the physical properties of the lens by combining the measurement of the parallax from the \Spitzer-``only" analysis with the measurement of $\theta_{\rm E} = 0.211 \pm 0.019 $ mas from fitting the ground-based light curve. These estimates and their uncertainties are derived from Equation (\ref{eqn:ML}) using simple error propagation, and so are the ``naive" values of these quantities. For the $(u_0 > 0)$ solution, this yields a lens mass of $M_{\rm L} = 0.073 \pm 0.011 M_{\odot}$ and $D_{\rm L} = 4.96 \pm 0.74$ kpc for $D_{\rm S} = 7.88$ kpc. This would then imply that the mass of the planet is $m_{\rm p} = 1.7 \pm 0.3\ M_{\oplus}$ and that it is separated from the host by $a_{\perp} = 1.01 \pm 0.18$ au. The values for the $(u_0 < 0)$ solution are comparable. See Table \ref{Table_bay}.

\begin{figure}[h]
\begin{center}
   \includegraphics[scale=0.55,angle=270]{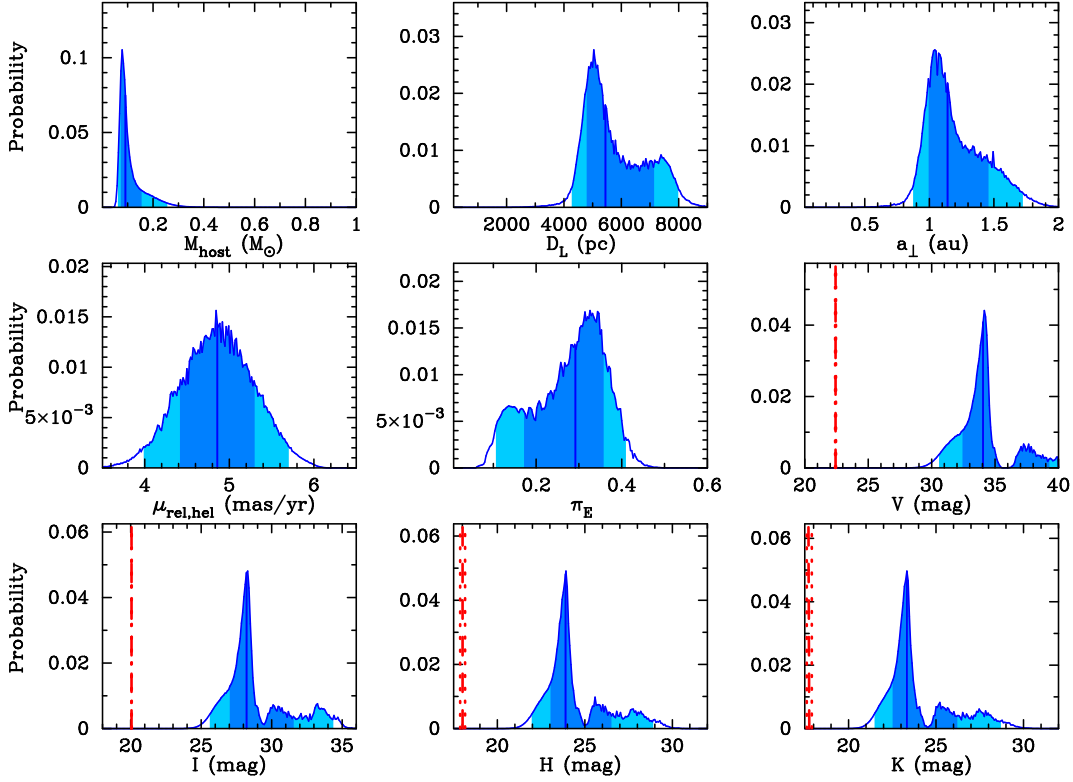}
\caption{Same as Figure \ref{Figure_bay_gr}, but with the addition of the $\pi_{\rm E}$ constraint from the $Spitzer$-``only'' parallax measurement.}
\end{center}
\label{Figure_bay_sp}
 \end{figure}

In order to estimate the lens magnitude, we also performed a Bayesian analysis including the $\pi_{\rm E}$ constraint derived from the $Spitzer$-``only'' parallax analysis. First, we took the average of the $\chi^2$ values for the two $(u_0 >0)$ and $(u_0 < 0)$ solutions for each value of $\pi_{\rm E,E}$ and $\pi_{\rm E,N}$.
Then, the event rate was weighted by $\exp(-{\Delta\chi^2}/2)$ and the measured $t_{\rm E}$ and $\theta_{\rm E}$ constraints to calculate the probability distribution. Table \ref{Table_bay} and Figure \ref{Figure_bay_sp} show the results. The distributions for some of the parameters in Figure \ref{Figure_bay_sp} are bimodal. In addition to the expected peak for lenses at $D_{\rm L} \sim 5$ kpc, there is a second peak for lenses with $D_{\rm L} \sim 7.5$ kpc. This second peak corresponds to events with lenses in the bulge and sources in the far-disk, which were not considered in our naive calculations. For the bimodal distributions, the central values and confidence intervals reported in Table \ref{Table_bay} are not a complete description of the distributions and should be considered in context with Figure \ref{Figure_bay_sp}. However, the mass distribution is not subject to this issue. We find that the lens system is likely a terrestrial planet with a mass of $m_{\rm p} = 2.1^{+1.5}_{-0.4}\ M_\oplus$ orbiting a very-low-mass (VLM) dwarf with a mass of $M_{\rm host} = 0.091^{+0.064}_{-0.018}\ M_\odot$.

\subsection{Implications}

\begin{figure}[h]
\begin{center}
	\includegraphics{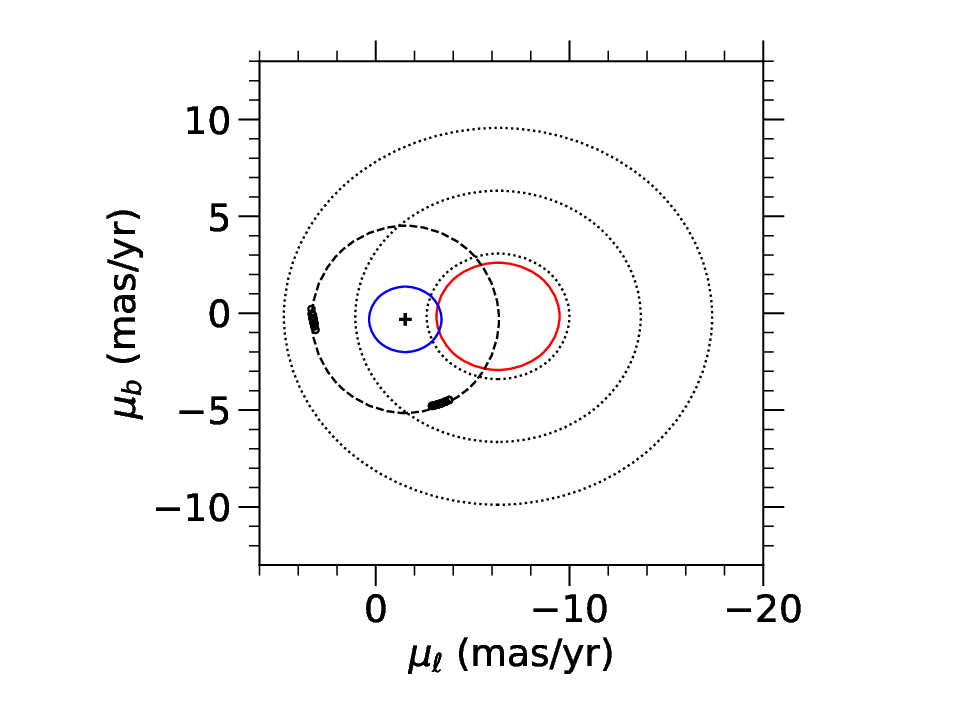}
	\caption{Test of source proper motion predicted by the \Spitzer-``only" parallax. {\em Black points}: derived source proper motions for $\pi_{\rm E}$ within $1\sigma$ of the minimum for the \Spitzer-``only" contours (based on $\mu_{\rm rel, hel}$). 
	 {\em Black cross}: mean proper motion for disk stars assuming a distance of $D_{\rm L} = 4.9\ $ kpc. {\em Dashed circle}: centered on black cross with a radius $\mu_{\rm rel, geo} = 4.832\ \mathrm{mas\ yr}^{-1}$.  Note that the black cross and dashed circle are merely reference points.
	{\em Red}: $1\sigma$ error ellipse for the bulge stars as derived from Gaia. {\em Blue}: $1\sigma$ error ellipse for the disk stars derived from $(\sigma_{v,\phi}, \sigma_{v,z})$. {\em Dotted black contours}: 1, 2, 3$\sigma$ contours adding the dispersions of the bulge and disk in quadrature. The observed constraints are consistent with a lens in the disk and a source in the bulge.\label{fig:muS}}
\end{center}
\end{figure}

Hence, if the \Spitzer-``only" parallax is correct, this would be the second detection of a terrestrial planet orbiting a VLM dwarf from the \Spitzer\ microlensing program. The first was OGLE-2016-BLG-1195Lb \citep{2017MNRAS.469.2434B,2017ApJ...840L...3S}, which is a $m_{\rm p} = {1.43}_{-0.32}^{+0.45}\ M_{\oplus}$ planet orbiting an $M_{\rm L} = 0.078_{-0.012}^{+0.016}\ M_{\odot}$ VLM dwarf at a separation of $a_{\perp} = 1.16_{-0.13}^{+0.16}$ au. The distance to the OGLE-2016-BLG-1195L system is also comparable: $D_{\rm L} = 3.91_{-0.46}^{+0.42}$ kpc. One curiosity about OGLE-2016-BLG-1195L is that the lens-source relative proper motion suggests that the lens could be moving counter to the direction of Galactic rotation, which would be unusual for a disk lens. 

Therefore, we also consider the implications of the \Spitzer-``only" $\bpi_{\rm E}$ for constraining the lens motion in \thisevent. First, we note that there is no independent information on the proper motion of the source $\mu_{\rm S}$ because there is no evidence that the blend, which dominates the baseline object, is associated with the event (see Appendix \ref{sec:appendix}). Second, given $D_{\rm L} \sim 4.9\ $ kpc, we assume that the lens is in the disk, and therefore, has a proper motion similar to other disk stars.
The velocity model of \citet{2021arXiv210403306K} is based on the Shu distribution function model in \citet{2014ApJ...793...51S}, but the mean velocity and velocity dispersion in the disk are fitted to the Gaia DR2 data \citep{GaiaDR2} as a function of the Galactocentric distance, $R$ and the height from the Galactic plane, $z$.
The velocity of disk stars at 4.9 kpc is $(v_\phi, v_z) = (207.6^{+42.7}_{-44.0}, -0.4^{+38.8}_{-39.6})\ \mathrm{km\ s}^{-1}$.
Hence, for the velocity dispersion, we use $(\sigma_{v,\phi}, \sigma_{v, z}) = (43.4, 39.2)\ \mathrm{km\ s}^{-1}$. Table \ref{tab-velo} summarizes the disk star velocities and proper motions expected from the Galactic model at $D=4.9\pm0.7$ kpc.
The values in the table are derived from the Bayesian analysis with a Galactic prior and constrained by $\theta_{\rm E}$ and $t_{\rm E}$. For the Sun's  motion, we use $(v_R, v_\phi, v_z)_{\rm Sun} = (-10, 243, 7)\ \mathrm{km\ s}^{-1}$ (for $(R_\odot, z_\odot) = (8160, 25)$ pc). We combine the two velocities to estimate the proper motion of disk stars.
Finally, by applying Equation \ref{eqn:muhel}, we can derive the expected source proper motion $\mu_{\rm S} = \mu_{\rm L} - \mu_{\rm rel, hel}$ for a given value of the parallax. Figure \ref{fig:muS} shows the results for values of $\bpi_{\rm E}$ out to the $1\sigma$ \Spitzer-``only" contours for the $(u > 0)$ solution (the results for the $(u < 0)$ solution are nearly identical). The properties of bulge stars are derived from Gaia stars within $5^{\prime}$ of the target: $\mu_{\rm bulge}(\ell, b) = (-6.310, -0.163) \pm (0.088,  0.076)\ \mathrm{mas\ yr}^{-1}$ and $\sigma_{\rm bulge}(\ell, b) = (3.176, 2.768) \pm (0.062,  0.054)\ \mathrm{mas\ yr}^{-1}$.
To account for the uncertainty in the lens motion, we add the proper motion dispersions of the disk and bulge in quadrature. One of the two \Spitzer\ minima suggests a source more than $2\sigma$ from the bulge distribution, but the other minimum is  consistent with a bulge source at $\sim 1.5\sigma$. Therefore, there is no reason to believe that the \Spitzer\ $\bpi_{\rm E}$ requires a lens proper motion in tension with the motion of typical disk stars.

 
 Finally, in order to be included in the statistical samples for the study of the Galactic distribution of planets, \citet[]{2017AJ....154..210Z} proposed the criteria:
\begin{equation}
\sigma(D_{8.3}) < 1.4 {\rm kpc};\ \ D_{8.3} \equiv \frac{\rm kpc}{\pi_{\rm rel}/{\rm mas} + 1/8.3}.
\end{equation}
We find $D_{8.3}=5.15 \pm 0.28$ kpc for the ($u_0>0$) case and $D_{8.3}=5.04 \pm 0.28$ kpc for the ($u_0<0$) case by combining the measurement of $\bpi_{\rm E}$ from the \Spitzer-``only" analysis with the measurement of $\theta_{\rm E}$ from fitting the ground-based light curve.
The small $\sigma(D_{8.3})$ is consistent with the expectation for the high magnification event as investigated by \citet{GouldYee12}, \citet{2018ApJ...863...23S}, and \citet{2019JKAS...52..121G}.
They show that accurate parallax measurements are possible even if there are only a few observations taken by the \Spitzer\ when the Earth-based magnification is high ($A_{\oplus} \geq 100$).
Therefore, in terms of $\sigma_{\rm D8.3}$ \citep[]{2017AJ....154..210Z}, the \Spitzer-``only" parallax suggests that the apparent signal is good enough to include \thisevent Lb in the statistical sample of \Spitzer\ events. However, the systematics need be studied and understood before membership in the sample can be definitively evaluated.

\begin{table}[h]
\caption{Disk star velocities and proper motions at $D=4.9 \pm 0.7$ kpc}
\begin{center}
\label{tab-velo}
 \begin{tabular}{c|cr|rrrrrr}
 \hline \hline
 Star Component & Velocity Component & Unit & $-2\sigma$ & $-1\sigma$ & Median& $+1\sigma$ & $+2\sigma$\\
 \hline
Thin Disk			&	$v_l$						& $\mathrm{km\ s}^{-1}$		&  110.7 & 163.6 &  205.9 &  242.4 &  280.7\\
				&	$v_b$						& $\mathrm{km\ s}^{-1}$		& -95.0 &  -48.8 & -13.7	& 22.4    & 71.1\\
				& 	$\mu_{{\rm hel},l}$			& $\mathrm{mas\ yr}^{-1}$		& -5.615 & -3.349& -1.577& -0.024 & 1.656 \\
				& 	$\mu_{{\rm hel},b}$			& $\mathrm{mas\ yr}^{-1}$		& -4.364 & -2.388&-0.884 & 0.653 & 2.690\\
 \hline
Thick Disk		&	$v_l$						& $\mathrm{km\ s}^{-1}$			& 60.7  & 125.2 &181.4 &236.5 &293.7 \\
				&	$v_b$						& $\mathrm{km\ s}^{-1}$			& -147.8& -86.4& -12.1& 63.3 & 128.6\\
				& 	$\mu_{{\rm hel},l}$			& $\mathrm{mas\ yr}^{-1}$	   	       & -7.662 & -4.995& -2.602& -0.275& 2.177 \\
				& 	$\mu_{{\rm hel},b}$			& $\mathrm{mas\ yr}^{-1}$ 		& -6.577 &-3.987 & -0.808 & 2.379 & 5.102\\
 \hline
All 				&	$v_l$						& $\mathrm{km\ s}^{-1}$			& 103.6 & 161.0 & 204.9 & 242.2&281.3 \\
				&	$v_b$						& $\mathrm{km\ s}^{-1}$			& -101.1 & -50.5 &-13.6 &24.4 & 77.4 \\
				& 	$\mu_{{\rm hel},l}$			&  $\mathrm{mas\ yr}^{-1}$      		& -5.878& -3.457& -1.620& -0.034& 1.685 \\
				& 	$\mu_{{\rm hel},b}$			&  $\mathrm{mas\ yr}^{-1}$		&-4.611 & -2.462& -0.883 & 0.737& 2.960\\
 \hline \hline
\end{tabular}
\end{center}
\end{table}


\section{Discussion and Summary}
\label{sec-dis}

We analyzed the microlensing event OGLE-2018-BLG-1185, which was simultaneously observed from a large number of ground-based telescopes and the $Spitzer$ telescope. 
The ground-based light-curve modeling indicates a small mass ratio of
$q = (6.9\pm 0.2) \times 10^{-5}$, which is close to the peak of the wide orbit exoplanet mass-ratio 
distribution derived by \citet{2016ApJ...833..145S} and investigated
further by  \citet{2018AcA....68....1U} and \citet{2019AJ....157...72J}.
\citet{2016ApJ...833..145S} derived the wide orbit planet occurrence rate using a sample of thirty planets, 
primarily from the MOA-II microlensing survey during $2007-2012$. 
The planet presented here, OGLE-2018-BLG-1185Lb,
will be included in an extension of the MOA-II statistical analysis (Suzuki et al., in preparation), and its low mass ratio will help to define the mass ratio function peak.

From the ground-based light-curve modeling, only finite source effect is detected, yielding a measurement of the angular Einstein radius. However, the physical properties of the lens as derived from the light curve are unclear because the observed flux variation of the $Spitzer$ light curve is marginal. Using only the constraint from the 
measured angular Einstein radius and a conservative constraint on the change in the \Spitzer\ flux, we estimate the host star and planet masses with a Bayesian analysis under the assumption that stars of all masses have an equal probability to host this planet. This analysis indicates 
a host mass of $M_{\rm host} = 0.37^{+0.35}_{-0.21}\ M_\odot$ and a planet mass of $m_p =  8.4^{+7.9}_{-4.7}\ M_\oplus$ located at $D_L = 7.4^{+0.5}_{-0.9}$ kpc.
By contrast, the $Spitzer$ data favor a larger microlensing parallax, which implies a very low-mass host with a terrestrial planet ($M_{\rm host} =  0.091^{+0.064}_{-0.018}\ M_\odot$, $m_{\rm p} = 2.1^{+1.5}_{-0.4}\ M_\oplus$) that is either in the disk at $D_{\rm L} \sim 5$ kpc or in the bulge at $D_{\rm L} \sim 7.5$ kpc (these values include a Galactic prior but are not significantly different from the values without the prior, see Table \ref{Table_bay}). 

\begin{figure}[htbp]
\begin{center}
   \includegraphics[scale=0.4,angle=0]{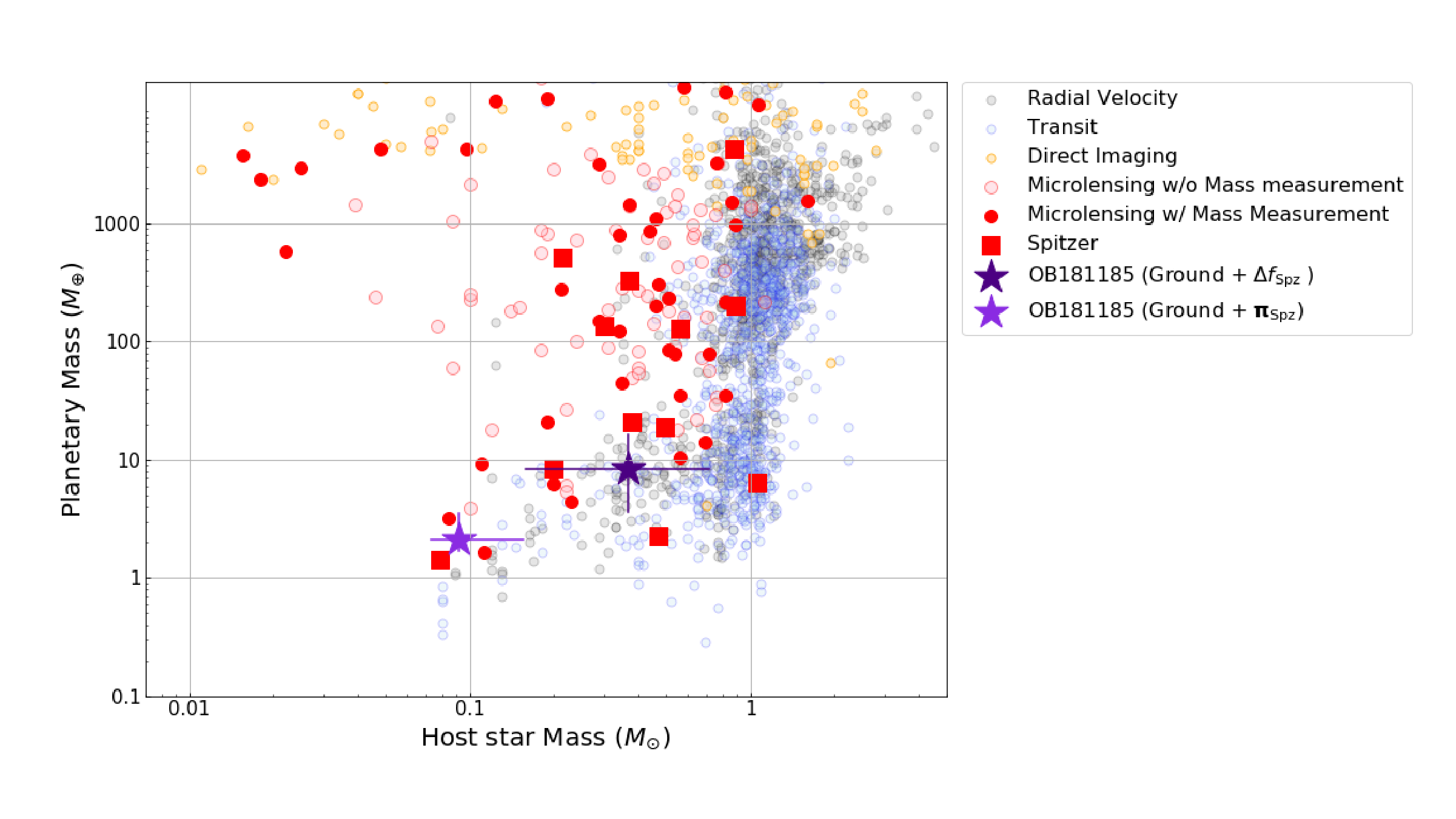}
\caption{The mass distribution of the detected exoplanets as of 2021 February 25 from http://exoplanetarchive.ipac.caltech.edu. The purple stars indicate OGLE-2018-BLG-1185. The pink circles show the microlens planets without mass measurements, and the red circles show the microlens planets with mass measurements from ground-based orbital parallax effects and/or the detection of the lens flux by the high resolution follow-up observations. The red squares represent the microlens planets with mass measurements from satellite parallax effects by $Spitzer$. The blue, yellow, and black dots indicate planets found by the transit, direct imaging, and radial velocity methods, respectively.} 
\end{center}
\label{Figure_distance}
 \end{figure}

Figure \ref{Figure_distance} compares the Bayesian estimates from the conservative \Spitzer\ flux constraint and the full \Spitzer\ parallax measurement of the host and planet mass for \thisevent\ to those of other planetary systems. The pink circles show the microlens planets without mass measurements, and the red circles show the microlens planets with mass measurements from ground-based orbital parallax effects and/or the detection of the lens flux by high resolution follow-up observations. The red squares represent microlens planets with mass measurements from the satellite parallax effect observed by $Spitzer$. Figure \ref{Figure_distance} indicates that if the \Spitzer\ parallax is correct, this is one of the lowest mass planets discovered by microlensing.

However, the result that this is a terrestrial planet orbiting a very-low-mass (VLM) dwarf in the disk should be treated with caution, because the amplitude of the \Spitzer\ signal is at the level of systematics seen in other events. A comparison of these properties to the Bayesian posteriors (Figure \ref{Figure_bay_gr}) demonstrates that a higher-mass system is preferred given $t_{\rm E}$, $\theta_{\rm E}$, and the Galactic priors. At the same time, a VLM-dwarf + terrestrial planet is still within the $2\sigma$ range of possibilities from the Bayesian analysis, especially once the constraint on $\Delta f_{\rm Spz}$ is imposed (Figure \ref{Figure_bayes_dflux}). Furthermore, 
\citet{2017ApJ...840L...3S} suggest that such planets might be common.  Nevertheless, further investigation is needed in order to assess whether or not the fitted parallax signal (and so the inferred mass) is real.

Adaptive optics observations are one way to test the \Spitzer\ parallax signal. The Bayesian analysis with ground-based + $\Delta f_{\rm Spz}$ constraints indicates the lens $K$-band magnitude with extinction should be $K= 20.8^{+1.8}_{-1.5}$ mag, which is about 3 magnitudes fainter than the source. By contrast, if the \Spitzer-``only" parallax is correct and the lens is a VLM dwarf, it should be $K = 23.3^{+2.9}_{-0.8}$ mag and therefore, much fainter and possibly undetectable. 
The Bayesian estimate of the heliocentric relative proper motion, $\mu_{{\rm rel, hel}} =  5.0 \pm 0.4$ mas ${\rm yr}^{-1}$, predicts that the angular separation between the source and the lens will be $\sim 30$ mas around mid-2024. Thus, the lens can be resolved from the source by the future follow-up observations with Keck or ELTs. If such resolved measurements were made (and the lens were luminous), it would also lead to a direct measurement of $\bmu$. The observed magnitude of $\bmu$ can serve as a check on $\theta_{\rm E}$. Additionally, the direction of $\bmu$ is the same as the direction of the microlens parallax vector, which could clarify how the \Spitzer-``only" parallax contours should be interpreted in the presence of systematics.

If the \Spitzer\ parallax is verified, this event confirms the potential of microlensing to measure the wide-orbit planet frequency into the terrestrial planet regime. Although the number of microlens planets with mass measurements is small for now, observing the satellite parallax effect can continue to increase the numbers. In particular, this effect can be measured for terrestrial planets by simultaneous observations between the ground and L2 \citep{GouldGaudiHan03}. This can be achieved with the PRIME telescope (PI: Takahiro, Sumi) and $Roman$ Space Telescope \citep{2015arXiv150303757S, 2019ApJS..241....3P} in the mid-2020s.

\software{OGLE\ \ DIA\ \ pipeline \citep{2003AcA....53..291U}, MOA\ \ DIA\ \ pipeline \citep{2001MNRAS.327..868B}, KMTNet\ \ pySIS\ \ pipeline \citep{2009MNRAS.397.2099A}, DanDIA \citep{2008MNRAS.386L..77B, 2013MNRAS.428.2275B}, DoPHOT \citep{1993PASP..105.1342S}, ISIS \citep{1998ApJ...503..325A, 2000A&AS..144..363A, 2018PASP..130j4401Z}, Image-centered\ \ ray-shooting\ \ method \citep{1996ApJ...472..660B, 2010ApJ...716.1408B}.}

\acknowledgements
Work by I.K. was supported by JSPS KAKENHI Grant Number 20J20633. 
Work by J.C.Y. was supported by JPL grant 1571564.
Work by D.P.B., A.B., and C.R. were supported by NASA through grant NASA-80NSSC18K0274. 
T.S. acknowledges the financial support from the JSPS, JSPS23103002, JSPS24253004, and JSPS26247023. 
Work by N.K. is supported by JSPS KAKENHI Grant Number JP18J00897. 
A.S. is a University of Auckland Doctoral Scholar. 
Y.T. acknowledges the support of DFG priority program SPP 1992 “Exploring the Diversity of Extrasolar Planets” (WA 1047/11-1). 
T.C.H acknowledges financial support from the National Research Foundation (NRF; No. 2019R1I1A1A01059609) 
U.G.J. acknowledges support from H2020-MSCA-ITN-2019, grant no.860470 (CHAMELEON) and the NovoNordisk Foundation grant no. NNF19OC0057374. 
W.Z. and S.M. acknowledge support by the National Science Foundation of China (Grant No. 11821303 and 11761131004). 
Work by C.H. was supported by the grants of National Research Foundation of Korea (2020R1A4A2002885 and 2019R1A2C2085965). 
Funding for B.S.G. was provided by NASA grant NNG16PJ32C and the Thomas Jefferson Chair for Discovery and Space Exploration.
The MOA project is supported by JSPS KAK-ENHI Grant Number JSPS24253004, JSPS26247023, JSPS23340064, JSPS15H00781, JP16H06287, 17H02871, and 19KK0082. 
The OGLE project has received funding from the National Science Centre, Poland, grant MAESTRO 2014/14/A/ST9/00121 to AU. 
This research has made use of the KMTNet system operated by the Korea Astronomy and Space Science Institute (KASI) and the data were obtained at three host sites of CTIO in Chile, SAAO in South Africa, and SSO in Australia. 
This research uses data obtained through the Telescope Access Program (TAP), which has been funded by the TAP member institutes. 
This work has made use of data from the European Space Agency (ESA) mission
{\it Gaia} (\url{https://www.cosmos.esa.int/gaia}), processed by the {\it Gaia}
Data Processing and Analysis Consortium (DPAC,
\url{https://www.cosmos.esa.int/web/gaia/dpac/consortium}). Funding for the DPAC
has been provided by national institutions, in particular the institutions
participating in the {\it Gaia} Multilateral Agreement. This research has made use of the NASA Exoplanet Archive, which is operated by the California Institute of Technology, under contract with the National Aeronautics and Space Administration under the Exoplanet Exploration Program.


{}

\appendix
{\section{Constraints on the Blended Light \& Discrepancy with {\em Gaia}}
\label{sec:appendix}}

The blended light in this event is roughly four times brighter than the source. In principle, the blend could be the lens itself or a companion to either the lens or the source. If so, it could constrain the flux and proper motion of the lens or the proper motion of the source.

From the KMTNet images, we measure the astrometric offset between the source and the baseline object and find an offset of $0\farcs175$. This offset is larger than the astrometric uncertainties. Therefore, if it is a companion to the lens or source, it must be a very wide separation companion ($\sim 1000\ {\rm au}$). However, the large separation also suggests that it could be an ambient star unrelated to the microlensing event. 

We measure the
proper motion of the baseline object based on 10 years of OGLE survey data and find
$\bmu_{\rm base}({\rm RA}, {\rm Dec}) = (-6.00 \pm 0.26, -4.25 \pm 0.16)\ \mathrm{mas\ yr}^{-1}$. Because the blend is much brighter than the source, its motion should dominate the measured $\bmu_{\rm base}$. The measured value is
very consistent with typical proper motions for normal bulge stars,
but not unreasonable for the proper motion of a disk star. Hence, it
does not rule out the possibility that the blend is a wide-separation
companion to the source or the lens, but it also shows that the blend
could easily be an unrelated bulge star. 

For completeness, we note that the OGLE measurement of the
proper motion of the baseline object is inconsistent with the reported
Gaia proper motion of the nearest Gaia source
\citep[4062756831332827136;][]{GaiaMission16}. Gaia EDR3 \citep{2020arXiv201201533G}
reports there is a G = 20.1 mag star $0\farcs177$ from the OGLE coordinates for the baseline star
(17:59:10.26 -27:50:06.3). The reported proper motion of this source
is $\bmu({\rm RA}, {\rm Dec}) = (-12.173 \pm 1.247, -9.714 \pm 0.870)\ \mathrm{mas\ yr}^{-1}$, which is an
outlier relative to the typical proper motions for stars in this
field. Gaia DR2 \citep{GaiaDR2} reports an only slightly less extreme
proper motion of $\bmu({\rm RA}, {\rm Dec}) = (-8.475 \pm 2.234, -4.039 \pm 1.985)\
\mathrm{mas\ yr}^{-1}$. The nature of this discrepancy is unknown, but because the Gaia proper motion is highly unusual (and the OGLE proper motion is typical), and the Gaia measurement varies significantly between DR2 and EDR3, this suggests a problem with the Gaia measurement.

\end{document}